\documentstyle[12pt,fleqn]{article}
\font\elevenbf=cmbx10 scaled\magstep 1
\renewcommand{\baselinestretch}{1.2}

\def\npb#1#2#3{    {\it Nucl. Phys. }{\bf B
#1}
(19#2) #3}
\def\plb#1#2#3{    {\it Phys. Lett. }{\bf B
#1}
(19#2) #3}
\def\prd#1#2#3{    {\it Phys. Rev. }{\bf D
#1}
(19#2) #3}
\def\prep#1#2#3{   {\it Phys. Rep. }{\bf #1}
(19#2)
#3}
\def\prl#1#2#3{    {\it Phys. Rev. Lett. }{\bf
#1}
(19#2) #3}

\def\ltap{\ \raisebox{-.4ex}{\rlap{$\sim$}}
\raisebox{.4ex}{$<$}\ }
\def\gtap{\ \raisebox{-.4ex}{\rlap{$\sim$}}
\raisebox{.4ex}{$>$}\ }

\def\abs#1{\left| #1\right|}

\def\dmk{{\Delta m_K}}
\def\dmB{{\Delta m_{B}}}

\def\tm{{\tilde m}}
\def\tq{{\tilde q}}
\def\refmark#1{\cite{#1}}

\def\tg{{\tilde g}}

\def\dI{{$\Delta I={1\over 2}~$}}
\def\B{{{\rm BR}_{\ell} (B) }}
\def\BF{{{\rm BR} (b \to s \gamma )}}
\def\BFd{{{\rm BR} (b \to d \gamma )}}
\def\g{{\gamma}}
\def\Qgs{{Q^{ds \pm }_{\scriptstyle G} }}
\def\Qgsp{{Q^{ds +}_{\scriptstyle G} }}
\def\Qgsm{{Q^{ds -}_{\scriptstyle G} }}
\def\Qgd{{Q^{db \pm }_{\scriptstyle G} }}

\def\Qgb{{Q^{sb \pm }_{\scriptstyle G} }}

\def\Qgdsm{{Q^{ds -}_{\scriptstyle G} }}

\def\cgs{{C_{\scriptstyle G}^{ds \pm} }}
\def\cgb{{C_{\scriptstyle G}^{sb \pm} }}
\def\cgd{{C_{\scriptstyle G}^{db \pm} }}

\def\cgsp{{C_{\scriptstyle G}^{ds +} }}
\def\cgsm{{C_{\scriptstyle G}^{ds -} }}
\def\cgbp{{C_{\scriptstyle G}^{sb +} }}
\def\cgbm{{C_{\scriptstyle G}^{sb -} }}
\def\cgdp{{C_{\scriptstyle G}^{db +} }}
\def\cgdm{{C_{\scriptstyle G}^{db -} }}

\def\cgds{{C_{\scriptstyle G}^{ds \pm} }}
\def\cgsb{{C_{\scriptstyle G}^{sb \pm} }}
\def\cgdb{{C_{\scriptstyle G}^{db \pm} }}

\def\cgdsp{{C_{\scriptstyle G}^{ds +} }}
\def\cgdsm{{C_{\scriptstyle G}^{ds -} }}

\def\Qfb{{Q^{sb \pm }_{\scriptstyle F} }}

\def\Qfd{{Q^{db \pm }_{\scriptstyle F} }}

\def\cfb{{C_{\scriptstyle F}^{sb \pm} }}
\def\cfbp{{C_{\scriptstyle F}^{sb +} }}
\def\cfbm{{C_{\scriptstyle F}^{sb -} }}
\def\cfd{{C_{\scriptstyle F}^{db \pm} }}
\def\cfdp{{C_{\scriptstyle F}^{db +} }}
\def\cfdm{{C_{\scriptstyle F}^{db -} }}

\def\dmdsp{{\scriptstyle \Delta} m_{ds}^+}
\def\dmdsm{{\scriptstyle \Delta} m_{ds}^-}
\def\dmds{{\scriptstyle \Delta}
m_{ds}^{\pm}}
\def\dmsbp{{\scriptstyle \Delta} m_{sb}^+}
\def\dmsbm{{\scriptstyle \Delta} m_{sb}^-}
\def\dmsb{\Delta m_{sb}^{\pm}}
\def\dmdbp{{\scriptstyle \Delta} m_{db}^+}
\def\dmdbm{{\scriptstyle \Delta} m_{db}^-}
\def\dmdb{{\scriptstyle \Delta}
m_{db}^{\pm}}

\def\a{{\alpha}}
\def\l{{\lambda}}
\def\b{{\beta}}
\def\d{{\delta}}
\def\D{{\scriptstyle \Delta}}
\def\om{{\omega}}

\def\e{{\epsilon}}
\def\0{{\over}}
\newcommand{\bea}{\begin{eqnarray}}
\newcommand{\beq}{\begin{equation}}
\newcommand{\eea}{\end{eqnarray}}
\newcommand{\eeq}{\end{equation}}

\newcommand{\spav}[1]{\parbox{1mm}{\vspace*{#1}}}

\begin{document}

\begin{titlepage}
\begin{flushright}
SLAC-PUB-6626\\
September, 1994\\
(T/E)
\end{flushright}
\spav{0.0cm}
\begin{center}
{\Large\bf Implications of TeV Flavor
Physics }\\
{\Large\bf for the \dI Rule and ${\rm
BR}_{\ell} (B)
$}\footnote{Work supported by the Department of
Energy, contract DE-AC03-76SF00515.} \\
\spav{1.4cm}\\
{\large Alexander L. Kagan}\footnote{Email: kagan@slac.stanford.edu,
Address after Sep. 1: Deptartment of Physics, University of Cincinnati.}
\spav{0.7cm}\\
{\em Stanford Linear Accelerator Center}\\
{\em Stanford University, Stanford,
California
94309}\\
\spav{1.0cm}\\

{\sc Abstract}
\end{center}

Two of the outstanding discrepancies between weak interaction
phenomenology and the standard model come in the large size of the
$\Delta I = {1\over  2}$ enhancement in $K$ decays and in the
small value of the $B$ semileptonic branching ratio.  We argue that
these discrepancies are naturally explained by chromomagnetic
dipole operators arising from new physics at the TeV scale.
These operators are closely connected to diagrams which contribute
to the quark mass matrix, and we show how the proper enhancement
of the hadronic decays  of $s$ and $b$ quarks can be linked to
generation of particular Cabbibo-Kobayaski-Maskawa mixing angles or
quark masses. We confirm our model-independent analysis with
detailed consideration of supersymmetric models and of technicolor models
with techniscalars. This picture has additional
phenomenological predictions for the $B$ system:  The branching ratio
of charmless nonleptonic $B$ decays should be of order 20\%, due to a
large rate for $b\to  sg$, while there
are no dangerous new contributions to $b\rightarrow s \gamma$.
Sizable contributions to $b \to d \gamma$ are a common feature of
models incorporating this mechanism. In techniscalar models
the $Z b \bar b$ coupling is enhanced, in association with sizable
contributions to $b \to s \mu^+ \mu^-$.

 \vfill

\end{titlepage}

\newpage
\setcounter{footnote}{0}
\setcounter{page}{1}

\vglue 0.2cm
{\elevenbf\noindent 1.  Introduction
}
\vglue 0.2cm

The \dI rule in $K \to \pi \pi $ decays
is one of the historical puzzles of particle physics.  The S-wave two pion
final
state
has
total isospin 0 or 2
and one has to understand why the \dI
transition
amplitude is
larger by a factor of twenty than the $\Delta
I={3\over 2}$
transition amplitude.
In the standard model a large
non-perturbative QCD
matrix-element enhancement is required.
Indeed, calculations
of the \dI amplitude
employing the ${1\over N_c}$ expansion and other models of strong interaction
behaviour at low energies give substantial enhancement
\refmark{bardeen,pichderafael,buras}.
Nevertheless, these estimates remain about a factor of
two too small after perturbative QCD corrections at
next-to-leading order are taken into account
\refmark{buras}.   In ref.
\refmark{stechneubert} a
phenomenological approach suggested that important contributions could come
from
effective diquark
states.
Final state
interactions might also enhance the \dI
amplitude
and suppress the $\Delta I={3\over 2}$
amplitude
\refmark{isgur}.
But neither approach is completely persuasive.
Twenty years after the birth of QCD, the large size of the \dI amplitude
remains
an
important puzzle.

In the $B$ system there is also persistent evidence for
discrepancy between existing measurements
and
the
standard model, in the semileptonic
branching ratio of
$B's$.
The world average \refmark{sharma} for $B$ mesons produced at the $\Upsilon
(4S)$ resonance
is
\beq \B = 10.29 \pm .06
\pm .27 \% ,\label{eq:cleo} \eeq
and for $B$ mesons produced at the $Z$ resonance it is
\beq \B = 11.33 \pm .22 \pm .41 .\label{eq:LEPbsl} \eeq
On the other hand, the parton model tends to
give
\refmark{ap} $\B  \gtap 13
\%$, including leading
\refmark{altarellimaiani,gaillardlee}
and next-to-leading
\refmark{altarellicleo,burasweisz}
order QCD
enhancement of the hadronic $B$ meson decay
width.
Again, one can appeal to non-perturbative effects to resolve the discrepancy.
However,
a recent analysis \refmark{bigi}
employing
heavy quark
effective field theory (HQET) techniques gave estimates of these
non-perturbative
terms which are much smaller than would be necessary.
(This is still controversial \refmark{falk}
and is sure to be debated further in the future.)

A possibly related anomaly may be present in measurements of the charm
multiplicity,
$n_c$, in $B$ decays.  Defined
as the number of charm states per $B$ decay, one obtains
$n_c \approx 1.2$ in the parton model.  This exceeds 1
because of the decay channel $b \to c \bar{c} s$.
Measured values have persistently exhibited a `charm deficit'.
Although the world average \refmark{muheim}
\beq n_c = 1.08 \pm .06 \label{eq:worldavgcharm} \eeq
has recently increased, it remains consistent
with a possible charm deficit.
This suggests that non-charm
hadronic decay channels are
somehow enhanced, thus simultaneously
suppressing $\B$.

Perhaps the data, together with improved
calculations of hadronic
flavor-changing
processes,
is starting to tell us something about a
possible role
for new flavor physics.
For example, it may turn out that the
standard
model \dI enhancement, while very large,
only
accounts for $50\%$ to $70\%$ of the
observed \dI
amplitude.   A significant portion of the \dI
rule
would still have to be
accounted for in this case.
It would be wonderful if the same mechanism
could give an additional, exotic, channel for
hadronic $B$ decays.

In this paper, I suggest the hypothesis that there are new
perturbative
contributions to $K$ and $B$ decay amplitudes
resulting
from chromomagnetic dipole operators
induced at TeV
energies.  These effects occur in a wide variety of models
with new interactions at the TeV scale.
For example, quark dipole moments are typical
in compositeness and extended technicolor scenarios.
In general, these new interactions are
closely connected to diagrams which contribute to the quark
mass matrix.
In particular, removal of
the gluon from a chromomagnetic dipole graph
often leaves a finite quark mass contribution.  Thus, $\B$
suppression and substantial
contributions to the \dI amplitude might
be the byproducts of new {\it flavor physics}
which also explains features of the
quark mass spectrum.
We will provide model-independent arguments as well as explicit
examples
which demonstrate that such a connection is
possible and, perhaps, even likely.

Some of the
earliest suggestions for the origin of the \dI
rule
\refmark{fritzchminkowski,ellis,vsz,hill}
involved new interactions which induce the
\dI chromomagnetic dipole operators
\beq \Qgs = g_s \overline{ d} \sigma_{\mu
\nu} t^a {{ 1 \pm \g_5
}\over 2} s G_a^{\mu \nu}
\label{eq:chromodipole12} \eeq
via penguin graphs.
In the standard model these operators are
suppressed by light quark masses and their
contribution to the \dI
amplitude is more than an order of
magnitude
smaller than the
conventional four-fermion operator
contributions.
However, the
requisite fermion chirality flip associated
with
dipole operators can
be much larger if these operators are induced by new
physics;
examples of
this have been presented from time to time in
the
literature.
In  \refmark{fritzchminkowski,ellis,vsz} the
\dI
chromomagnetic dipole operators
were induced via charm-changing right-handed
charged currents
coupled to the $W$ boson.
Of course this possibility has long since been ruled out.
In \refmark{hill} these operators were discussed in the context of
multi-Higgs doublet models but their
contributions
were suppressed
by light quark masses.
An $E_6$ inspired model was
considered in
\refmark{ma} in which the \dI chromomagnetic
operators were
generated via loop diagrams containing vectorlike down
quarks
and
neutral scalars.
Finally, the authors of \refmark{tracas},
again
motivated by $E_6$, found that scalar
diquark exchange could generate a substantial \dI
amplitude via the chromomagnetic dipole operaors.
The
authors of refs.
\refmark{ma,tracas} eventually reached
negative
conclusions
after invoking constraints on their models
from $K-\overline{K}$ mixing.

We will demonstrate here that the \dI chromomagnetic
dipole operators
can acquire large
coefficients in supersymmetric models, and in
technicolor
models which employ techniscalars to generate quark masses.
The corresponding contributions to
the (\dI) $K
\to \pi \pi $
amplitude first arise at order
$p^4 $ in the chiral
lagrangian expansion \refmark{holstein}, and are
unfortunately difficult to estimate.
But they could well
account for $30\% - 50\%$ of the
observed amplitude, which would significantly narrow
any gap
between
theory and
experiment.
We also find that for supersymmetry with ultra-light
gluinos
the induced \dI amplitude can be larger.
In all of our
examples we check that
the most stringent
constraints on flavor-changing neutral
currents are not violated.
Our results are, in
particular, consistent with the known small
value of
$ \Delta m_K \equiv m(K_L) - m(K_S) $.

New physics can also induce the $\Delta B=1 $
chromomagnetic dipole
operators
\beq \Qgb = g_s \overline{ s}
\sigma_{\mu
\nu} t^a {{1 \pm \g_5}\over 2} b G_a^{\mu \nu}
\label{eq:chromodipole23} \eeq
\beq \Qgd = g_s \overline{ d} \sigma_{\mu
\nu} t^a
{{1 \pm \g_5
}\over 2} b G_a^{\mu \nu}
\label{eq:chromodipole13} \eeq
with significantly larger coefficients than in
the
standard
model.
The resulting enhancement of the $B$ meson
hadronic
decay width
could be large enough to solve the $\B$
puzzle.  The operators in (\ref{eq:chromodipole23})
can increase the branching ratio for $b
\to s g$ to  $15-30 \%$, well above what is expected in the standard
model.
This possibility
was first pointed out in
\refmark{hou}
in the context of
two-Higgs doublet models,
and
more recently
in \refmark{bigi}.
This type of resolution would also lead to a
charm
deficit in $B$ decays which is consistent
with the measured value.

It is important to check that models of $\B$ suppression do not
produce large unwanted contributions to ${\rm BR}(b \to s \gamma)$.
CLEO has recently announced a measurement of this branching ratio
\refmark{cleoBF},
\beq {\rm BR} (B \to s \gamma) = (2.32 \pm .51 \pm .29 \pm.32) \times 10^{-4} ,
\label{eq:Bsgamma} \eeq
which corresponds to an upper bound of $4 \times 10^{-4}$.
Unfortunately, multi-Higgs doublet models of
$\B$ suppression are in gross conflict with this bound \refmark{hounew}
and are thus excluded.\footnote{The possibility of dangerously large
contributions, in general, in multi-Higgs doublet models has
been discussed in refs. \refmark{hewett,barger}.}
A simple model-independent criterion will be introduced which can be used to
identify
models of $\B$
suppression which do not run into
this difficulty.

We will see that in
supersymmetric models
and in
technicolor models with techniscalars
it is
easy to
induce large enough coefficients for the
$\Delta B
=1$ chromomagnetic
dipole operators
to resolve the $\B$ puzzle.
$B- \bar B$ mixing constraints are not
restrictive, and
electric-dipole
contributions to $b \to s \gamma $ are
sufficiently
small.  In certain cases $\BFd$ is 1 to 2 orders of magnitude larger
than in the standard model, lying in the range
$(.1 -1.0)\times 10^{-4}$.  This has
interesting implications for
observation of
$B \to \rho \gamma$ or $B \to \phi \gamma$ at CLEO and future $B$
factories.  The present bound \refmark{sharma} is
\beq {\rm BR}(B \to \rho \gamma)
< .34 \cdot {\rm BR}(B \to K^* \gamma) \label{eq:Brhogamma}
\eeq
at $90\%$ CL, which leaves a large window open for new physics.

The main point of this paper is to uncover a possible connection
between certain features of the quark mass spectrum and the various
puzzles outlined above.
Our model-independent analysis will suggest that
a substantial portion of the \dI
amplitude is directly associated
with mass contributions which account for
$m_s$, or
$\theta_c$.  The analysis also suggests that
resolutions of the $\B$
puzzle
attributed to chromomagnetic dipole
operators are directly associated with generation of $m_b$,
with $\sim 100~MeV$ mass contributions which
account for
$V_{cb}$ (and $m_s$), or with
smaller
mass contributions which account for
$V_{ub}$ (as well as $\theta_c$ and $m_d$).
The supersymmetry and technicolor examples will
illustrate these points explicitly.  Two phenomenologically distinct
possibilities for a new
scale of flavor physics
emerge:  $M \sim 1-2~TeV$ ({\it{Region I}}), which can be associated with $\B$
suppression, and
$M \sim {1 \over 2 }~TeV$
({\it{Region II}}), which can be associated with both $\B$ suppression and \dI
enhancement.

\vglue 0.3cm

We organize our discussion as follows.  To
further
motivate the
introduction of new
physics we begin in Section 2 with a review
of the
status of the \dI
rule and inclusive $B$ decays in the standard
model.
A model-independent discussion of the
phenomenology of chromomagnetic and
electric-dipole operators and associated quark mass contributions
follows in Section 3.
Supersymmetry and technicolor examples are
discussed in
Sections 4 and 5, respectively.  A discussion and summary of our results is
given in
Section 6.
Appendix A provides further details on the
relationship between the dipole operators and the
quark mass spectrum, and Appendix B contains
expressions for
new
contributions to
$\dmk$ and $\dmB$ in the models we consider.

\vglue 0.6cm
{\elevenbf\noindent 2.  The \dI rule and
$\B$ in
the standard model
}
\vglue 0.2cm

The amplitudes for $K^0 \to \pi^+ \pi^-$ and
$K^0
\to \pi^0 \pi^0$ can be
parametrized in terms of the \dI transition
amplitude, $A_0$, and
the $\Delta I ={3\over 2 }$ transition
amplitude,
$A_2$, defined as
\beq  A_I  = \langle (\pi \pi )_ I \vert H_W
\vert
K^0
\rangle,~~~I=0,2 . \label{eq:AI} \eeq
$H_W$ is the weak hamiltonian and the
subscripts
0, 2 denote the
total $\pi \pi$ isospin.  Experimentally
\refmark{A0A2exp}
\beq \abs{A_0} = 3.3 \times 10^{-
7}~GeV,~~~~~\abs{A_2 }= 1.5
\times 10^{-8}~GeV, \label{eq:A0A2} \eeq
and the \dI rule is manifested by the ratio
$\abs{A_0 /  A_2 }= 22.2$.

In the standard model the bulk of the \dI
amplitude is almost certainly due to the
4-quark operator hamiltonian
\beq H_W = C_1 Q_1 + C_2 Q_2 ,\label{eq:HW} \eeq
where
\bea Q_1 & = & [\overline{s}_{\a} \g_{\mu}
(1-\g_5
)  d_{\b} ]
[\overline{u}_{\b} \g^{\mu} (1-\g_5)  d_{\a} ]
\nonumber \\
Q_2 & = & [\overline{s} \g_{\mu} (1-
\g_5 )  d][ \overline{u} \g^{\mu}  (1-\g_5)  d]
.\label{eq:Q12} \eea
The \dI matrix elements can be expressed as
\refmark{buras}
\bea \langle (\pi\pi)_0
\vert  Q_1 \vert K^0 \rangle & = & -{1\over
9}\sqrt{3\over
2}F_{\pi}(m_K^2
-m_{\pi}^2) B_1^{(1/2)} \nonumber \\
\langle (\pi\pi)_0
\vert  Q_2 \vert K^0 \rangle & = & {5\over
9}\sqrt{3\over 2}F_{\pi}(m_K^2 -m_{\pi}^2)
B_2^{(1/2)},
\label{eq:Q1Q2} \eea
where $F_\pi = 132~MeV$.
The parameters $B_1^{(1/2)}$ and
$B_2^{(1/2)}$
are both equal to 1
in
the vacuum insertion approximation.  In the
${1\over N_c}$
approximation they are
enhanced \refmark{bardeen} to
approximately
$5.2$ and
$2.2$, respectively, at $\mu \approx
.6~GeV$, which
corresponds to
$B_2^{(1/2)}  (m_c) \approx 2.8$.
Qualitatively similar conclusions have been
reached
in refs. \refmark{pichderafael,jaminpich}.

A naive estimate of the resulting \dI
amplitude at
zero'th order in QCD,
\beq A_0^{V-A} \sim {G_F \over
\sqrt{2}}
V_{ud} V_{us}^* \langle
(\pi\pi)_0
\vert  Q_2 \vert K^0 \rangle,
\eeq
falls an order of magnitude short of
experiment in the
vacuum insertion approximation and a factor of 3
short in the ${1\over N_c}$ approximation.
The authors of
\refmark{buras} find, taking leading and
next-to-leading order
QCD corrections of the Wilson coefficients and
matrix-elements in (\ref{eq:HW})
into
account, that phenomenologically building
the  \dI
amplitude
into the standard model requires
$B_2^{(1/2)}(m_c)
\sim 6$. This is about a factor of 2
larger than obtained in
the
${1 \over N_c}$ approximation \refmark{bardeen}, suggesting that
there might be new contributions to the \dI amplitude.

Next we summarize the status of the $B$ meson
semileptonic branching
ratio
in the standard model following the
parton
model analysis of ref. \refmark{ap} and the
recent
discussion of ref.
\refmark{bigi}.  The semileptonic decay
width of
$B$ mesons
in the parton model is given to ${\cal O}
(\alpha_s)
$ by
\beq  \Gamma_{\ell} = \Gamma (b \to c {\ell}
\overline{\nu}_{\ell})
=
\Gamma_0  I_0 ({m_c^2 \over
m_b^2},{m_{\ell}^2 \over
m_b^2},0)
[1-{{2 \a_s} \over{ 3 \pi} } f({m_c^2 \over
m_b^2},{m_{\ell}^2 \over m_b^2})
+{\cal O} (\a_s^2)],
\label{eq:smslwidth} \eeq
where
\beq  ~~~~~~\Gamma _0 \equiv {{G_F ^2 m_b^5
\abs{V_{bc}}^2 }\over
{192
\pi^3}},
 \nonumber \eeq
and $m_b$ is the pole mass.  Expressions for the phase space factor
$I_0$
for
negligible electron or
muon mass or non-negligible $\tau$ mass in the
final
state can be found
in ref. \refmark{cortes}.  The function $f$ is
given
explicitly in ref.
\refmark{sirlin} and has been tabulated in
ref.
\refmark{maianicabibbo}.

There are two classes of non-leptonic decays.
For
down and strange
quarks in the final state one obtains
\beq  \Gamma (b \to c \overline{u}d) +
\Gamma (b
\to c
\overline{u}s)
= 3 \Gamma_0 I_0 ({m_c^2 \over m_b^2},0,0)
\eta J
\label{eq:smhadwidth}
\eeq
For $\Gamma (b \to c \bar{c}s)$ one obtains
an
analogous expression
with
the substitution
\beq I_0 ({m_c^2 \over m_b^2},0,0) \to I_0
({m_c^2 \over
m_b^2},0,{m_c^2
\over
m_b^2}) \nonumber \eeq
In eq. (\ref{eq:smhadwidth}) $\eta$ is the leading-log anomalous
dimension
enhancement
\refmark{altarellimaiani,gaillardlee} and
$J$ is the
enhancement
due to next-to-leading corrections
\refmark{altarellicleo,burasweisz}.
The total branching ratio for charmless $b$ decays in the standard model is
expected to be
$1-2 \%$.  We omit these decays from our discussion
since they have a negligible effect on $\B$ and $n_c$ for our purposes.

The expected value of the semileptonic branching ratio depends strongly
on
$m_b$, $m_c$ and
$\Lambda_{QCD} $.
Varying \refmark{hou} $m_b$ and $m_c$
independently, the lowest
value for $\B$ is obtained for maximal
$m_b$ and
minimal $m_c$;
keeping $m_b -m_c $ fixed
$\B$ increases with increasing $m_b$.  In
the
parton model the
electron spectrum in $B$ decays implies
\refmark{ruckl}
$m_b -m_c =3.37 \pm .03~GeV$, which is in good agreement with
the difference obtained in HQET.
The
authors of ref. \refmark{nir} have found that
the $B$
semileptonic
decay rates imply $m_b \ge 4.66~GeV$ and
$m_c
\ge 1.43 ~GeV$ in HQET.
Finally, recent lattice
calculations
\refmark{langnau} give  $m_b =4.94 \pm .15~GeV$.

In Fig. 1 we plot parton model predictions
for
$\B$ versus $\alpha_s (M_Z) $ at
the
renormalization point\footnote{For
the expressions
used in \refmark{ap} $\eta J$ is only
$\mu$ independent to order $\alpha_s$,
reflecting
our ignorance of
order $\alpha_s^2$
corrections.  It was noted that for $\mu =
{m_b /
2}$ the QCD
corrections are enhanced and one approaches
the
observed values of
$\B$.  This is, however, an extreme
possibility.} $\mu =
m_b$.
We have checked that our plot is in good agreement with
ref.
\refmark{ap}
for $m_b =4.6~GeV$ and $m_c =1.2~GeV$.
For less extreme
choices of $m_b$
and $m_c$ one clearly expects $\B > 12.5
\%$.  As
an illustration,
for
$m_b =4.8~GeV$ and $m_c = 1.4~GeV$ we
obtain $\B=13.4 \%, (13\%)$
for
$\Lambda_{QCD}^{(4)} = 300~MeV,(400~MeV)$.
This is to be contrasted with the measured values,
which are subsantially lower.

The authors of
\refmark{bigi}
have
estimated non-perturbative  ${ \cal O} \left( {1 / m_b^2 } \right)$ and
higher-order corrections to
the parton model approximation in the heavy quark expansion
and
find a very small
decrease, $\delta \B \sim -.3 \%$.
Of course it may turn out that the operator product expansion fails
for non-leptonic decays \refmark{falk}. Although the total energy
released is much larger than $\Lambda_{QCD}$,
the energy per strongly interacting particle is considerably smaller
than in semileptonic decays. This is especially relevant in the two
charm decay channel where resonance effects may become important in the
final hadronic state. However, the parton model $b \to c \bar{c} s$ decay rate
would
have to be doubled in order to obtain agreement with
measurements of $\B$.  Of course, another possibility is that
$\B$ suppression is due to a combination of
non-perturbative QCD effects and new physics.

Finally, we discuss the
expected charm multiplicity for $B$ decays in the parton
model. The amount by which $n_c$ exceeds 1 is identified with $BR
(b \to c
\bar{c} s)$.  For example, for $m_b
=4.8~GeV$, $m_c =1.4~GeV$, $\mu =m_b$ and $\alpha_s (M_Z) \approx .11-.13$,
we obtain
$n_c \approx  1.2$.
This essentially agrees with the heavy quark expansion results of ref.
\refmark{stechpalmer},
in which $n_c =1.19 \pm .01$ is obtained for $m_b =4.8~GeV$, $m_c =1.35~GeV$,
and
$\alpha_s (M_W) =.12$.
Lower values of $m_c$, while decreasing $\B$ will
increase $n_c$ slightly.\footnote{For example, $m_c
=1.2~GeV$ and $m_b =4.6~GeV$ gives $n_c
\approx
1.25$.}
As already
noted, the measured
multiplicity
is consistent with a small `charm
deficit'.  This would appear to rule out enhancement of the $b \to c \bar{c} s
$
rate as the
origin of $\B$ suppression, and instead
suggests that there are sizable new contributions to charmless $b$ decays.

\vglue 0.6cm
{\elevenbf\noindent 3.  Phenomenology of the
Quark Dipole Operators, and the Quark Mass Spectrum}
\vglue 0.2cm

This section is devoted to a model-independent
discussion of the
phenomenology of dipole penguin operators induced
by new {\it flavor physics}
in the
context of the \dI rule and
$\B$.  In particular, we will determine what
ranges
for the operator
coefficients correspond to significant
enhancements
of the \dI
amplitude and the $B$ meson hadronic decay
width.  This,
in turn, will
have implications for the scale of new flavor physics which induces these
diagrams, and
for the associated induced quark masses and
additional flavor-changing
effects.  The
operator
coefficients and
induced masses are taken real throughout.
We briefly remark on CP violation
in the
Conclusion.

We begin with discussion of the chromomagnetic
dipole operators
defined in eqs. (\ref{eq:chromodipole12}), (\ref{eq:chromodipole23}), and
(\ref{eq:chromodipole13}).  Contributions of
electromagnetic dipole operators to
radiative $B$ meson
decays are
discussed later.
The relevant terms in the chromomagnetic dipole Lagrangian are
shown explicitly below
\beq { \Delta}{\cal
L}_{\scriptstyle G} =
\sum _{i=+,-}
C_{\scriptstyle G}^{ds  i} (\mu)
Q_{\scriptstyle
G}^{ds i} (\mu) +
\sum _{i=+,-} C_{\scriptstyle G}^{sb i}
(\mu)
Q_{\scriptstyle G}^{sb
i}(\mu) +\sum _{i=+,-} C_{\scriptstyle
G}^{db i}
(\mu) Q_{\scriptstyle G}^{db
i}(\mu)  +H.c. +...\label{eq:lagrangian} \eeq
The $C_{\scriptstyle G}$ are operator
coefficients of
dimension
$(mass)^{-1}$.
At leading
order in QCD their evolution obeys the
relation
\refmark{vsz2,ellis,hill}
\beq C_{\scriptstyle G}^{\pm} (\mu_2) =
\left({
{\alpha_s (\mu_2 )}
\over \alpha_s
(\mu_1) } \right)^{-{2 \over {3b}}}
C_{\scriptstyle
G}^{\pm} (\mu_1)
,\label{eq:cglowenergy} \eeq
where $b= 11 -2 n_g - {2/3} n_f$. $n_f$ is
the
number of flavors and
$n_g$ is the number of gluinos
( 0 or 1).\footnote{In supersymmetric models we
will identify the scale of new physics with the squark
masses and take the gluinos lighter than the squarks.  We therefore
do not include squark contributions to the $\beta$ functions.}
The
small
anomalous
dimension leads to a small decrease in the coefficients
of about $10\%$ when evolving from
TeV scales to the $b$ scale.
Unless otherwise specified, we use the
following
numerical inputs
and thresholds for
evolution of operator coefficients:
$\Lambda_{QCD}
^{(4)}
=300~MeV$,
$m_t =170~GeV$, $m_b = 4.8~GeV$, and
$m_c =1.4~GeV$.

The operator coefficients $\cgsp$, $\cgbp$
and
$\cgdp$ are also additively
renormalized due to mixing with the
standard
model dimension-six
operators at ${\cal O} (\alpha_s^2)$
\refmark{admixing}.
The largest effect, due to mixing with $Q_2$,
changes $\cgsp$ by ${\cal O} (3\% )$
and $\cgbp$ by ${\cal O} (10\%)$, if these coefficients
have magnitudes in the ranges of interest for
\dI enhancement and $\B$ suppression.  The
relative sign of these
contributions is not fixed.
In the interest of simplicity we
consider
only the multiplicative renormalization and
ignore these additional small
corrections.

\vglue 0.3cm
\noindent {\bf \it   Parametrization of Flavor Physics}
\vglue 0.2cm

In general, each dipole operator coefficient might
receive several
new contributions.
In the following analysis we parametrize the
case in which there is a single source for all of the coefficients,
corresponding to a single
exchange of particles.  Some examples are
exchange of a single gluino-squark pair at
one loop in supersymmetric theories,
exchange of a single techniboson-technifermion pair
in technicolor theories, and exchange of a
quark-charged scalar pair in multi-Higgs doublet models.
For a single exchange, the induced operator coefficient
matrix and the induced quark mass matrix
are proportional and of unit rank.
It is straightforward to generalize
to the case of several contributions, leading to
matrices of rank 2 or 3.

We will deal with two quark basis,
the quark mass eigenstates, or physical quarks, as usual denoted
by  $d_L$, $s_L$, etc., and the interaction basis quarks, which are denoted by
$d_L^i$,
$d_R^i$, $i =1,2,3$.
New interactions generate dipole coefficient matrices and quark mass matrices
in
the
interaction basis.
The physical transition dipole moments
are obtained by taking matrix elements of these matrices in the mass
eigenstate basis.
In general, in the quark interaction basis
we write ${ \Delta}{\cal
L}_{\scriptstyle G}$ as
\beq { \Delta}{\cal
L}_{\scriptstyle G} =
\sum _{i,j}
C_{\scriptstyle G}^{ij} (\mu)
Q_{\scriptstyle
G}^{ij} (\mu) + H.c. , \label{eq:Lginteraction} \eeq
where
\beq Q_{\scriptstyle
G}^{ij} = g_s \overline{d}^ i_{\scriptstyle L} \sigma_{\mu
\nu} t^a d^j_{\scriptstyle R}  G_a^{\mu \nu}.
\label{eq:Qginteraction} \eeq
The corresponding mass contributions, obtained by
removing the gluons
from the dipole graphs, are
\beq { \Delta} {\cal L}_{mass} =
\sum_{i,j} {\scriptstyle \Delta}
m_{ij}
\bar{d}^i_L d^j_R ~+~H.c. \label{eq:Lminteraction} \eeq

Restricting to the case of
a single exchange of particles, we can parametrize the
coefficients in the following way
\beq C_{\scriptstyle G}^{ij} (\mu)  =  \eta (\mu)
\zeta_{\scriptstyle G}
{{{\scriptstyle \Delta}
m_{ij}
 (M) } \over
M^2} . \label{eq:cijmij} \eeq
As noted above, $C_{\scriptstyle G}^{ij}$ and
${\scriptstyle \Delta}
m_{ij}$ will be proportional rank 1 matrices.
$M$ is the scale of new physics,
identified with the mass of the
heaviest particle exchanged.
In the supersymmetric examples it will be
identified with
the mass of the exchanged squark
and in the technicolor examples it will be
identified with the
mass of the
exchanged techniscalar.
$\eta$ is a dimensionless parameter which accounts for
multiplicative renormalization from $M$ down to hadronic
mass scales, as discussed above.
All of the flavor information is contained in
the
induced quark
masses.  The remaining model dependence is then represented by the flavor
independent
and $\mu$-independent parameter $\zeta_{\scriptstyle G}$.
Simple dimensional analysis reveals that the
dipole operator coefficient must be ${\cal O}
\left({{\scriptstyle
\Delta} m} \over
M^2 \right)$, so that $\zeta_{\scriptstyle G}$ is nominally
${\cal O} (1)$.
Indeed, in
the supersymmetric case
$\zeta_{\scriptstyle G}$
typically varies
between $1\over 2$ and 2, depending on the
squark and
gluino masses which enter the loop integrals,
while
in the technicolor case
it is approximately $1\over 2$.

The induced quark masses
associated
with the transition dipole operators in eq. (\ref{eq:lagrangian})
are obtained by taking matrix elements of
${\scriptstyle
\Delta} m_{ij} $ in the
quark mass eigenstate basis.  These are written as
\bea {\Delta} {\cal L'}_{mass} &
= &
{\scriptstyle \Delta}
m_{ds}^+
\bar{d}_L s_R ~+ ~{\scriptstyle \Delta}
m_{ds}^-
\bar{d}_R s_L~+~{\scriptstyle \Delta}
m_{sb}^+
\bar{s}_L b_R \nonumber \\
& + & {\scriptstyle \Delta}
m_{sb}^-
\bar{s}_R b_L
+~{\scriptstyle \Delta}
m_{db}^+ \bar{d}_L b_R~+~{\scriptstyle \Delta}
m_{db}^-
\bar{d}_R b_L~+~H.c.
\label{eq:masslagrangian}
\eea
The physical dipole operator coefficients
are given in terms
of these masses by
\beq \cgb (\mu)   =  \eta (\mu)
\zeta_{\scriptstyle G}  {{{\scriptstyle \Delta
}
m_{sb}^{\pm}  (M)
}\over
M^2}, ~~~~\cgd (\mu)   =  \eta (\mu)
\zeta_{\scriptstyle G}  {{{\scriptstyle \Delta
}
m_{db}^{\pm}  (M)
}\over
M^2} \eeq
\beq ~~~~~~~~~~~~~~~~~\cgs (\mu)   =  \eta (\mu)
\zeta_{\scriptstyle G}
{{{\scriptstyle \Delta }
m_{ds}^{\pm}  (M) }\over
M^2}. \label{eq:parametrization} \eeq
Given several contributions to the dipole operator
coefficients each of them can be parametrized as above,
although in general
$M$ and $\zeta_{\scriptstyle G}$ will differ in each case.

The ranges for the induced quark masses in (\ref{eq:parametrization})
which would
strongly suggest a
connection to the observed quark mass spectrum are
found
by expressing these masses
in terms of the interaction basis entries, ${\scriptstyle \Delta }m_{ij}$.
This is straightforward given reasonably
general assumptions about the hierarchy obeyed by
entries of the full down
quark mass matrix in the interaction
basis.\footnote{There will be some
uncertainty due to possible  cancelations among different
sources of quark mass and between up and
down
sector
contributions to the KM angles.}  Details are
provided in Appendix A.  Given the hierarchy of
eq. (A1) one concludes the
following:\footnote{The numbers in parenthesis,
evaluated at $\mu =m_c $, are
illustrative and
correspond to the running masses $m_c (m_c) =1.4~GeV$,
$m_b (m_c) = 5.4~GeV$ (or $m_b (m_b)
\approx 4.25~GeV$),
$m_s
(m_c) =150~MeV$, and $V_{cb} = .043$, $\abs{ V_{ub}/V_{cb}}
=.1$.}
\vglue 0.2cm
\noindent {\bf(a)}  If
$\abs{{\scriptstyle \Delta } m_{ds}^{+} }
\sim \abs{\theta_c m_s}$ ($\sim 33~MeV$) then the induced
unit-rank mass matrix, ${\scriptstyle \D} m_{ij} $,
can be associated with
generation of the bulk of $\theta_c $ or
$m_s$, but not both.
\vglue 0.2cm
\noindent {\bf(b)}  If $\abs{{\scriptstyle \Delta } m_{sb}^{+} }
\sim \abs{V_{cb} m_b }$ ($\sim 230~MeV$) then
${\scriptstyle \D} m_{ij} $
can be associated with generation of the
bulk of
$V_{cb}$ or $m_b$, but not both.
\vglue 0.2cm
\noindent{\bf(c)}  If $\abs{{\scriptstyle \Delta } m_{db}^{+} }
\sim \abs{V_{ub} m_b }$ ($\sim 23~MeV$) then
${\scriptstyle \D} m_{ij} $
can, in general, be associated
with generation of the bulk of
$V_{ub}$, $V_{cb}$, or $m_b$, but not all three.
\vglue 0.2cm

The question of
which features of the quark mass
spectrum are
in fact generated in (a)-(c) above
is a model-dependent issue which we address when
discussing specific examples.
In principle, all of the KM
angles and down quark masses can be associated with induced
dipole operator coefficients
given several sources for these operators.

With the above parametrization we can study
\dI enhancement and $\B$ suppression due to new
flavor physics in a model-independent way in the appropriate
($\zeta_{\scriptstyle G}
{\scriptstyle \D} m$,$~M$) plane.
Two model-independent conditions constrain the
allowed regions
of \dI anhancement and $\B$ suppression in these planes:
\vglue 0.2cm
\noindent {\bf(i)} The scale of new flavor physics should lie
above the
weak scale in order to have avoided detection.
\vglue 0.2cm
\noindent {\bf(ii)} The induced quark masses
should not spoil the observed quark
mass hierarchy.  Since $\zeta_{\scriptstyle G}$
is nominally of ${\cal O} (1)$ this means that
$\zeta_{\scriptstyle G}
{\scriptstyle \D} m$ should not be much larger than the
corresponding range
in (a)-(c) above in order to avoid fine-tuning of the
quark mass spectrum.\footnote{For example,
the amount of tuning
of $\theta_c $ or $m_s $ associated with the magnitude
of $\dmdsp$ is of order one part in $\abs{\dmdsp \over {\theta_c
m_s} }$.}
\vglue 0.2cm

We will see that $\B$ suppression and
\dI enhancement of a reasonable magnitude can be obtained with $\zeta_G \sim 1$
and
induced masses in the ranges specified in (a)-(c) above.  This implies that, in
general,
a connection with the observed quark
mass spectrum is possible.
Specific models can be classified according to where they
lie in the planes of ${\scriptstyle \D}m$ vs. $M$, or according to whether such
a
connection can be realized.
Flavor-changing constraints will rule out parts of the planes
and one has to make sure that the models survive these restrictions.

\vglue 0.3cm
\noindent{ \bf    The $B$ Meson Semileptonic Branching
Ratio}
\vglue 0.2cm

We begin with discussion of $\B$.  Estimates
in the
$B$ system are
more reliable and easier to present.  The
parton
model
contribution of the dipole operators $\Qgb $
to the inclusive hadronic decay width of $B$ mesons is
given by
\beq \Gamma (b \to sg)  =  {4 \over 3}
\alpha_s
(m_b) m_b^3
\left( \abs{\cgbp
(m_b)}^2 + \abs{\cgbm  (m_b)}^2 \right).
\label{eq:hadronicdecay1}
\eeq
In terms of our parametrization this is
\beq \Gamma (b \to sg) =   {4 \over 3}
\eta ^2 (m_b) \alpha_s (m_b)  m_b^5
\abs{V_{cb}}^2 {\zeta^2_{\scriptstyle G}
\over M^4}
({\scriptstyle \Delta } {m_{sb}^+}(M) ^2 +
{\scriptstyle \Delta } {m_{sb}^-}(M)^2)
 .   \label{eq:hadronicdecay2} \eeq
Expressions for the contribution of $\Qgd$
to
$\Gamma (b \to dg)$
are analogous, with $s$ indices
replaced everywhere
by $d$ indices.  The inclusive gluon channel decay width,
$\Gamma(b \to x g)$, is proportional
to
\beq \cgbp ^2 +\cgbm ^2
+\cgdp^2 + \cgdm
^2 \label{eq:cgp} \eeq
evaluated at $m_b$.
If there is only one source or exchange of particles
giving rise to the dipole operators then it is also proportional
to
\beq {{\scriptstyle \D} m' }^2 \equiv
{\dmsbp}^2
+{\dmsbm}^2+{\dmdbp}^2 +{\dmdbm}^2 ,\label{eq:dmell} \eeq
 evaluated at $M$.

In order to study the connection to the quark mass
spectrum it is convenient
to parametrize ${\scriptstyle \D} m'$ as
\beq {\scriptstyle \D} m' (\mu)= \xi'
\abs{V_{cb} m_b (\mu)}, \label{eq:xiell}
 \eeq
where $\xi' $ is a $\mu$-independent dimensionless parameter.
For illustrative purposes we assume that
$\abs{\cgbm} \ltap \abs {\cgbp}$,  $\abs{\cgdm} \ltap \abs {\cgdp}$
or, equivalently, that
$\abs{\dmsbm} \ltap \abs {\dmsbp}$, $\abs{\dmdbm} \ltap \abs
{\dmdbp}$.\footnote{If instead $\dmsbm >> \dmsbp$
and $\dmdbm >> \dmdbp$, which does not necessarily spoil the
quark mass spectrum, then the scale of new physics associated
with $\B$ suppresion is increased
but the connection to the KM matrix is lost.}
According to our previous discussion there are then two regions of
interest for $\xi'$ (or ${\scriptstyle \Delta} m'$):

\vglue 0.2cm

{\noindent\bf Case (I)} $\xi' \sim 1$ (or
${\scriptstyle \D}
m' \sim \abs{V_{cb} m_b }$), taken together with
the small $V_{ub}$ to $V_{cb} $ ratio,
suggests
the  hierarchy
\beq \abs{{\scriptstyle \Delta } {m_{sb}^+} }\sim \abs{
V_{cb} m_b },~~~~\abs{{\scriptstyle \Delta } {m_{db}^+}}
\sim \abs{V_{ub} m_b }. \label{eq:xione} \eeq
So $\xi' \sim 1$ can be associated with generation of $V_{cb}$
or $m_b$.

\vglue 0.2cm

{\noindent\bf Case (II)} $\xi' \sim .1$ (or ${\scriptstyle \D}
m' \sim \abs{V_{ub} m_b }$) is consistent with
\beq \abs{{\scriptstyle \Delta } {m_{db}^+}}
\sim \abs{V_{ub} m_b },~~~~\abs{{\scriptstyle \Delta } {m_{sb}^+}
}\sim\abs{
V_{ub} m_b }. \label{eq:xitenth} \eeq
So $\xi' \sim .1$ can be associated with generation of $V_{ub}$.
Alternatively, $\xi' \sim .1$ can be associated with generation\footnote{In
this
case the induced
mass matrix is assumed to account
for the bulk of
$m_{23}^d $ and $m_{33}^d$, leading to suppression of $\dmsbp$.}
of $V_{cb}$ in conjunction with $m_b$. Details are given in Appendix A, see eq.
(A6).
\vglue 0.2cm

In Fig. 2 we plot contours for $\B = 10\%$
and $11
\%$
in the ($\abs{\zeta_{\scriptstyle G} \xi'} $, $M$) plane.
$\eta (m_b) $ has been obtained with non-
supersymmertic $\beta $-
functions but it is nearly the same in
supersymmetric models. Note that large
uncertainties in $\B$
due to lack of
precise
knowledge of $V_{cb}$ and $m_b$
conveniently drop out in this
parametrization since
the gluon channel decay width is proportional to
$m^5_b
V_{cb}^2$, like the standard model decay
widths.
The parton
model charm
multiplicities for Fig. 2 are
$n_c =.9$ ($\B=10\%$) and
$n_c
=1.0$ ($\B=11\%$).  The latter is in better
agreement with
the measured charm multiplicity than the standard model
prediction.  The inclusive gluon channel branching ratios
are ${\rm BR}(b\to x g) = 25\%$ ($\B=10\%$) and $18\%$ ($\B=11\%$),
about an order of magnitude above the standard model prediction.\footnote{If
$\B$
suppression is due to a combination of new physics and non-perturbative
enhancement of
$\Gamma (b \to c \bar{c} s)$ then $n_c$ would be increased and ${\rm BR}( b \to
xg)$
would be decreased.  For example, keeping $\B$ fixed at $11\%$, a $20\%$
enhancement
of the two charm decay rate would shift $n_c$ by $\approx +.05$ and ${\rm BR}(
b
\to
xg)$ by $\approx - 7\%$.  However, our conclusions concerning quark mass
generation
would not change qualitatively.}

{}From Fig. 2 it is clear that the desired $\B$ suppression
can, in principle, take place in
either region of $\xi'$ of
relevance to the
quark mass spectrum.
\vglue 0.2cm
\noindent{$\bullet$} Case (I)
corresponds to a
scale of new physics $M \sim 1-2~TeV$.
Henceforth, we refer to this scale
of new physics as {\bf{Region I}}.
In this case $\Gamma (b \to dg) << \Gamma (b \to sg) $.
\vglue 0.2cm
\noindent{$\bullet$} Case (II)
corresponds to a somewhat lower scale of
new physics $M \sim 300-700~GeV$.
Henceforth, we refer to this scale of
new physics as {\bf{Region II}}. In this
case $\Gamma (b \to dg) \ltap \Gamma (b \to sg)$
is possible.

We will discuss
specific examples of new flavor physics which
feature considerable overlap with
one or the other region of the (${\scriptstyle \D} m' , M$) plane.
But first we discuss what is potentially the most restrictive
flavor-changing constraint associated with $B$ hadronic decay
enhancement.

\vglue 0.3cm
\noindent{ \it   $b
\to s
\gamma $ and $b \to d \gamma $}
\vglue 0.2cm

In general,
enhancement of the $B$ meson hadronic decay
width
will be correlated with contributions to $BR(b
\to s
\gamma)$ or ${\rm BR}(b \to d \gamma)$ due to the
induced electromagnetic dipole
operators
\bea \Qfb & =  & e Q_d \overline{s }
\sigma_{\mu \nu}  {{1 \pm \gamma_5 }
\over 2} b F^{\mu \nu}  \nonumber \\
\Qfd & =  & e Q_d \overline{d }
\sigma_{\mu \nu}  {{1 \pm \gamma_5 }
\over 2} b F^{\mu \nu}
,\label{eq:electricdipole } \eea
where $Q_d$ is the electric charge of the
down
quark.  An important question is whether the
hadronic enhancement associated with $\B$ is
consistent with the CLEO bound on the inclusive radiative branching ratio,
the sum
of $\BF$ and
$\BFd$.
The answer is model-dependent and we will
give a
general criterion which can be used to distinguish those models
in which the contribution of new physics
is not too
large.  On the other hand, $B_d - \overline{B}_d $
mixing
constraints are not restrictive, as will
become clear
when we discuss specific examples.

The Lagrangian for electromagnetic dipole
operators is
\beq  { \Delta}{\cal
L}_{\scriptstyle F} =
 C_{\scriptstyle F}^{sb+}
Q_{\scriptstyle
F}^{sb
+} +
 C_{\scriptstyle F}^{sb -}
Q_{\scriptstyle
F}^{sb
-} + C_{\scriptstyle F}^{db+} Q_{\scriptstyle F}^{db
+}+
 C_{\scriptstyle F}^{db -}
Q_{\scriptstyle
F}^{db
-} + H.c. \label{eq:lagrangianF} \eeq
Note that in general
the relative sign between new physics contributions to
$\cfbp$ and the standard model
contribution to $\cfbp$ is not fixed and
they can interfere
destructively or constructively.
In attempting to determine which models
do not give dangerously large
contributions to the inclusive radiative
branching ratio we can ignore the standard model contribution.
This will not
alter our conclusions qualitatively.

At leading-order, renormalization of the
operator
coefficients is given by
\refmark{admixing}
\beq C_{\scriptstyle F}^{\pm} (\mu_2) =
\left(
{{\alpha_s (\mu_2 )}
\over \alpha_s
(\mu_1) }\right)^{-{4 \over {3b}}}
C_{\scriptstyle
F}^{\pm} (\mu_1)
+ 2 \left[\left( {{\alpha_s (\mu_2 )}
\over \alpha_s
(\mu_1) }\right)^{-{4 \over {3b}}}-\left(
{{\alpha_s
(\mu_2 )}
\over \alpha_s
(\mu_1) }\right)^{-{2 \over {3b}}}\right]
C_{\scriptstyle G}^{\pm} (\mu_1).
\label{eq:cflowenergy} \eeq
Again we ignore mixing with the standard model four-fermion
operator
$Q_2 $ since the resulting contribution
to ${\rm BR}(b \to s \gamma)$ is essentially the same as in the
standard
model.
The relative sign between
$C_{\scriptstyle F}^{+}$ and
$C_{\scriptstyle G}^{+}$
, or $C_{\scriptstyle F}^{-}$ and
$C_{\scriptstyle
G}^{-}$ is model dependent and
renormalization due
to mixing with the chromomagnetic dipole
operators can be constructive or destructive.

Applying our parametrization for a single source
for dipole operators
to the electromagnetic dipole operator
coefficients gives
\beq \cfb (M)   =  \zeta_{\scriptstyle F}
{{{\scriptstyle \Delta }
m_{sb}^{\pm}  (M) }\over
M^2} ~~~GeV^{-1}, ~~~~\cfd (M)   =
\zeta_{\scriptstyle F}  {{{\scriptstyle \Delta
}
m_{sb}^{\pm}  (M)
}\over
M^2} ~~~GeV^{-
1}.\label{eq:parametrizationF} \eeq
$\zeta_{\scriptstyle F} $ is a
dimensionless, $\mu$-independent parameter
which is the analog of $\zeta_{\scriptstyle G}$
for photon emmision.
Again, all of the flavor dependence is
contained in
the induced quark masses.

The
inclusive decay width for $b \to s \gamma$
is given by
\beq \Gamma (b \to s \gamma)  =
\alpha_{em} Q_d^2
m_b^3
(\abs{\cfbp (m_b)}^2 + \abs{\cfbm (m_b)}^2)
{}.
\label{eq:photonicdecay}
\eeq
$\Gamma (b \to d \gamma)$ is
analogous,
with $s$ indices again replaced by $d$ indices.
The total $B$ radiative decay width is proportional to
\beq
\cfbp  ^2 + \cfbm  ^2 +\cfdp
 ^2 + \cfdm
  ^2 , \label{eq:cfp} \eeq
evaluated at $m_b$.

To arrive at a model-independent criterion which
insures that
the radiative branching ratio will not be too large
we need to determine what is a sufficiently small
magnitude for the ratio
$\zeta _{\scriptstyle F} \over \zeta_{\scriptstyle G} $,
given that $\Gamma (b \to x g)$ gives the desired
$\B$ suppression.
In Fig. 3 we plot this ratio for several representative values of $\B$ and
${\rm BR}(b \to x \gamma) $.\footnote{We have evolved the chromomagnetic and
electromagnetic dipole operator coefficients from $m_b$ to $M$ in order to
determine this
ratio.}
We have used non-supersymmetric
$\beta$-functions above $m_t$, but the supersymmetric case is
nearly
the same, again exhibiting a weak scale dependence.
Results have been included for $\zeta_F$ and $\zeta_G$ of same, or
opposite sign.

The ratio ${\zeta_{\scriptstyle F} \over  \zeta_{\scriptstyle G}
}$ is a model-dependent quantity which, in general,
will depend on the charges of
the particles which radiate the photon, ratios of
loop integrals, etc.
Essentially, what we find from Fig. 3 is that
{\it models
which give the desired $\B$ suppression
should
satisfy}
\beq \abs{\zeta_{\scriptstyle F} }<
\abs{\zeta_{\scriptstyle G}}
\label{eq:BFcriterion}\eeq
in order to insure that new contributions to ${\rm BR}(b\to x\gamma)$ are
sufficiently
small.
It is important to realize that
$\BFd$ provides a
very large window for new physics, since it is
two orders of magnitude smaller than
$\BF$ in the standard model.  In fact,
in those models in which
$\B$ suppression takes place in Region II,
$\BFd \sim (.1-1) \times 10^{-4}$ is
likely
since $\dmdb$ and $\dmsb$ tend to be of same order,
see eq. (\ref{eq:xitenth}).

If the chromomagnetic and electromagnetic dipole operators are due
to
gluon and
photon emmission from the same particle of charge
$Q$,
then ${\zeta_{\scriptstyle F} \over
\zeta_{\scriptstyle G} } = {Q\over Q_d}$, where $Q_d = -{1 \over 3}$.  This
implies that
in multi-Higgs doublet models of $\B$ suppression,
$\zeta_{\scriptstyle
F} $ is larger in magnitude than
$\zeta_{\scriptstyle G} $ because the dominant loop integral for
photon emmission corresponds to radiation from the charge ${2
\over 3}$ top quark.  Therefore, one can not
simultaneously obtain the desired hadronic
width
enhancement and
satisfy the CLEO bound.  Similarly, one can rule out
$E_6$-motivated models of $\B$ suppression in which the chromomagnetic dipole
operators are due to penguin graphs with scalar diquarks and a top quark in the
loop.
The case of dipole penguin graphs with neutral scalars and
vectorlike quarks
in the loop is borderline.
Since the photon and
gluon
are both emmitted from charge $-
{1 \over 3}$
vectorlike down quarks, $\zeta_{\scriptstyle
F} =
\zeta_{\scriptstyle G} $ and modest destructive
interference with the standard model
penguin contribution would
be required. Although potentially interesting, we will not discuss this
model further.

In the supersymmetric examples which we consider,
a gluino and squark are exchanged at one loop.
$\zeta_{\scriptstyle F}$ will be smaller in magnitude than
$\zeta_{\scriptstyle G} $ because the loop
integral
for photon emmission, corresponding to
emmission from the squark, is
significantly smaller than the dominant loop integral for
gluon
emmission, corresponding to emmision from the gluino.  In the
technicolor
examples which we consider,
$\abs{\zeta_{\scriptstyle F} \over \zeta_{\scriptstyle G}
} \approx  {1 \over 2}$ because both the photon and gluon are
emmitted
from a techniscalar
with
charge ${1 \over 6}$, or $\abs{Q_d}\over 2 $.
We
will see explicitly in Sections 4 and 5 that
$b \to s
\gamma $ and $b \to d \gamma$ constraints
are not very
restrictive in these examples.

\vglue 0.4cm
\noindent{\bf  The \dI Amplitude}
\vglue 0.2cm

As already noted, it is difficult to estimate
the
$K^0 \to \pi \pi $ amplitude induced by the
dipole
operators $\Qgs$.
The lowest order representation of $\Qgs$ in the chiral
lagrangian vanishes due to an exact cancelation at leading order in chiral
perturbation
theory between the direct
$K \to \pi \pi $ amplitude
and a pole contribution combining the strong
interaction $KK \pi \pi
$ vertex and
the $K$-vacuum tadpole \refmark{holstein,cheng}.  This can be
seen directly by using PCAC soft pion theorems to relate
the $K \to \pi \pi $
and $K \to $ vacuum matrix elements of $\Qgs$.
The reason for this cancelation is that the lowest
order representation is similar in form to the mass
term in the strong interaction lagrangian.
As a result, it can be rotated away by a chiral transformation without
inducing any other $\Delta S = 1$ terms in the lagrangian.

The leading-order chiral representation of
$\Delta {\cal L}_{\scriptstyle G}$ for
$K \to \pi \pi$ decay arises at ${\cal O} (p^4) $
and is of the form \refmark{holstein}
\beq {a \over \Lambda^2} Tr[\lambda_6 U
\partial_{\mu}
U^{\dagger}
\partial^{\mu} U ] +H.c.,\label{eq:nextterm} \eeq
where $\Lambda$ is of order the
chiral
symmetry breaking scale,
$\Lambda_{\chi SB}$.
Following ref.
\refmark{holstein} we make a crude estimate of the resulting \dI
amplitude by assuming that it is suppressed by ${p^2 \over
\Lambda^2} \approx {m_{K}^2 \over \Lambda^2} $
relative to the `direct' PCAC $K \to \pi \pi $ amplitude.
We write it as
\beq A_0 = ( C_{\scriptstyle G}^{ds +} (\mu)-
C_{\scriptstyle G}^{ds -}
(\mu) )
  \langle (\pi \pi)_{0} \vert \Qgsp (\mu)
\vert K^0
\rangle {m_K ^2
\over
\Lambda ^2} .\label{eq:dIamplitude} \eeq
Although we expect $ \Lambda_{\chi SB} \sim 1~GeV$
\refmark{manohar,donoghue}, $\Lambda $
can vary
substantially, in general, for higher order chiral lagrangian
contributions,
depending on which process or diagram is being considered.
The suppression factor in eq. (\ref{eq:dIamplitude})
could, a priori,
lie
anywhere in the
interval
${m_K ^2 \over \Lambda ^2} \sim .1 -.4$.  This is certainly
the case for higher-order contributions to the
\dI rule in the standard
model \refmark{bijnens,kambor}.
We will therefore present all of our results for
the \dI amplitude in terms of ${m_K ^2 \over \Lambda ^2}$,
keeping it as a phenomenological parameter to be
determined in the future.

We use a PCAC calculation
\refmark{matrixelement} of the `direct' matrix element in
eq. (\ref{eq:dIamplitude}), which gives
\beq \langle (\pi \pi)_{0} \vert \Qgsp \vert
K^0
\rangle = -
\sqrt{{3\over 2}} {m_0^2 \over 2 }
{m_K^2 \over {m_u + m_s }} {F_K \over
F_{\pi} ^2 }
B_{\scriptstyle
G}^{(1/2)}.
\label{eq:QG}\eeq
The decay constants are $F_ {\pi} = 132~MeV
$ and
$F_K =
161~MeV$.
$m_0^2$ parametrizes the
relevant mixed
condensate,
\beq g_s \langle 0 \vert \overline {q} \sigma
_{\mu \nu } T^a G_a
^{\mu \nu}
q \vert 0 \rangle = m_0^2 \langle 0 \vert
\overline
{q} q \vert 0
\rangle,
\nonumber \eeq
and $B_{\scriptstyle G}^{(1/2)} $ is a
dimensionless
matrix element
parameter
which is approximately equal to 1.\footnote{The
authors of
ref.
\refmark{matrixelement} take
$B_{\scriptstyle
G}^{(1/2)} (m_c) = 1$, based on the assumption that it is reasonable
to evaluate the matrix element at $m_c$.}
The two most recent determinations of
$m_0^2$ are
a lattice
calculation \refmark{kremer} and
a fit using QCD sum rules and $B$-meson
data\refmark{narison}, which give $m_0^2 (m_c)
\approx 1~GeV^2$ for $m_c =1.4~GeV$, or
\beq \langle (\pi \pi)_{0} \vert \Qgsp
\vert K^0
\rangle \approx
-9~GeV^2 \label{eq:QGnumber} \eeq
for $m_s (m_c) =150~MeV$.  We will make use of this result
throughout
and evaluate the operator coefficients in eq. (\ref{eq:dIamplitude})
at $m_c$.

In order to uncover a possible connection between \dI
amplitude enhancement
and generation of $\theta_c$ or $m_s$ it is useful
to parametrize the induced masses in eq.
(\ref{eq:parametrization}) as
\beq  {\scriptstyle \Delta } m_{ds}^{\pm}
(\mu)  =
\xi _{ds}^{\pm}  \abs{  \theta_c m_s (\mu)}
\label{eq:mdsparametrization}, \eeq
where, as usual, $\xi _{ds}^{\pm}$ are dimensionless
$\mu$-independent
parameters.  According to our previous discussion of
induced masses,
generation of $\theta_c$ or $m_s$ would correspond to
$\xi_{ds}^+ \sim 1$.

In terms of our parametrization, the \dI
amplitude
is given by
\beq A_0 = {{\zeta_{\scriptstyle G} (\xi_{ds}^+ -\xi_{ds}^-)
\eta (m_c)
\theta_c m_s (M)} \over
M^2}  \langle (\pi \pi)_{0} \vert \Qgsp
\vert K^0
\rangle   {m_K ^2
\over
\Lambda ^2}. \label{eq:dIparametrization} \eeq
It is important to point out
that
comparison of the observed \dI and $\Delta I = {3 \over 2}$
amplitudes in $K \to 3 \pi $ decays and $K
\to 2
\pi$ decays constrains the chiral
structure of the \dI amplitude
\refmark{golowich}.  In particular,
current algebra relations imply that
the contribution of
$\Qgsm$ to the $K
\to \pi\pi$
amplitude\footnote{I would like to thank John Donoghue
for bringing
this point and ref. \refmark{golowich} to my
attention.} should be small, perhaps $\ltap 10\%$. Equivalently, if
the dipole operators account for $30\%-50\%$ of the
\dI amplitude then
\beq \abs{\dmdsp} \gtap (2 - 4)  \abs{\dmdsm}, \label{eq:K3pi} \eeq
or $\abs{\xi_{ds}^+} \gtap (2-4)\abs{\xi_{ds}^-}$
should be satisfied.

In Fig. 4 we plot
contours of constant $R_0$, defined as the
ratio of magnitudes
of the dipole induced \dI amplitude, $A_0$, to
the observed \dI amplitude $A_0^{exp} $,
\beq R_0 \equiv \abs{A_0 \over A_0^{exp} } ,
\label{eq:r0} \eeq
in the plane of
$\abs{ \zeta_{\scriptstyle
G} (\xi_{ds}^+ -\xi_{ds}^-) }$ vs. $M$.
To first
approximation, the vertical axis in Fig. 4 can be
identified with $\abs{\zeta_{\scriptstyle G} \xi_{ds}^+ }$
for large
\dI enhancements.
Again, $\eta$ is nearly the same in supersymmetric models.  For purposes of
comparison we have also reproduced
contours of $\B$
from Fig. 2 in the ($\abs{\zeta_{\scriptstyle G} \xi'}$,$M$) plane.

Our model-independent analysis reveals that
$\zeta_{\scriptstyle G} {\scriptstyle \Delta }
 m_{ds}^{+} \approx
\theta_c m_s $ together with $R_0
\approx (1-1.5) {m^2_K \over  \Lambda
^2 }$ can be obtained in
the $M \sim {1 \over
2}~TeV$ region, identified as Region II in
our discussion of $\B$.
In general, we expect
$\zeta_{\scriptstyle G}\sim 1 $ so that
we can associate this region with
generation of $\theta_c $ or $m_s$.
If ${m^2_K / \Lambda ^2 }$ lies in the range $.2-.4 $,
generation of $ 30 \%-60 \% $ of the observed
\dI
amplitude is possible.
Since the relative sign between
the standard model 4-quark operator contribution and new
dipole operator contributions to the \dI
amplitude is generally not fixed, the two could add
constructively
helping to account for the entire
\dI amplitude.

In the next two sections we will discuss
supersymmetric and techniscalar models.
In particular, we will see that in both cases
substantial overlap with the above region of Fig. 4 is not
ruled out by the small value of $m (K_L) -
m(K_S) $, although in the supersymmetric examples
a modest one part in two or three tuning may be required
for the larger \dI amplitudes.
Gluinos in the `light-gluino' window will constitute a special
case. Because of the
extreme ratio of gluino to squark masses
entering the relevant loop integral,
$\zeta_{\scriptstyle G}$
will be substantially larger than 1.
The induced quark masses will generally be too
small to be of significance,
but very large \dI amplitudes will be possible for
squark masses below $500~GeV$.

\vglue 0.4cm

To summarize, we have performed a
model-independent analysis of
potential contributions of chromomagnetic dipole operators to the
$B$ hadronic decay width and the (\dI) $K \to \pi \pi $ amplitude.
By comparing results for \dI enhancement
and $\B$
suppression in Fig. 4 we can loosely identify two
interesting
scales of new physics, or $M$.
\vglue 0.2cm

\noindent{$\bullet$} In {\it{Region I}}, corresponding to $M \sim {1-2} ~TeV$,
the desired $\B$ suppression can be directly associated with
generation of $V_{cb}$ or $m_b$ (but not both).
However, substantial \dI enhancement would lead to
undesirably large contributions to $\theta_c $ or $m_s$.
\vglue 0.2cm
\noindent{$\bullet$} In {\it{Region II}}, corresponding to $M \sim {1\over 2}
{}~TeV$,
the desired $\B$ suppression can be directly associated
with generation of $V_{ub}$, or with
generation of $V_{cb}$ in conjunction with $m_b$.
The magnitude of the induced \dI amplitude is difficult
to estimate.  However, it can be as large as $30\%$ to $60\%$
of the observed
amplitude, without resorting to unreasonably large matrix elements.
Furthermore, it can be
directly associated with generation of  $\theta_c$ or $m_s$.
The question of
which masses or mixing angles are actually generated in
Region I or Region II is
model-dependent.
\vglue 0.2cm

\noindent{$\bullet$} Finally, we have given a general
criterion which can be used to distinguish
those models of $\B$
suppression in which the branching ratios for
$b\to s \gamma$ and
$b \to d \gamma$
are not too large.  We have also argued that in Region II
it is possible to obtain $\BFd \approx 10^{-4}$,
a dramatic departure from the standard model prediction.

We are now ready to discuss models
which illustrate the above points explicitly.

\vglue 0.6cm
{\elevenbf\noindent 4.  Supersymmetry}
\vglue 0.2cm

In this section we will discuss the phenomenology of radiatively
induced dipole operators in supersymmetric models.
We begin by setting some notation.
Superpartners are denoted by tildes.  For example, the gluino mass is
$m_{\tilde g}$.
Left-handed and right-handed down squarks are denoted by
$\tilde{d}^i_L$, $\tilde{d}^i_R$,
$i=1,2,3$ in the quark interaction basis, and by
$\tilde{d}_L$, $\tilde{d}_R$,
$\tilde{s}_L$, $\tilde{s}_R$, $\tilde{b}_L$, $\tilde{b}_R$
in the physical quark basis.
We make the usual assumption of an approximately
degenerate or universal flavor-diagonal squark
masses, $m_{\tilde q}^2$, corresponding to the following terms
in the squark mass matrix,
\beq \sum_{i=1,2,3} m_{\tilde q}^2 (\tilde{d}^{i^*}_L \tilde{d}^i_L
+ \tilde{d}^{i^*}_R \tilde{d}^i_R  ) .\label{eq:univsquarkmasses} \eeq

Deviations from universality are of two types.
Additional non-universal left-left and right-right squark
masses
\beq \sum _{ij} \d \tm^2_{i_L j_L} \tilde{d}^{i^*}_L  \tilde{d}^j_L +
\sum_{ij} \d \tm^2_{i_R j_R} \tilde{d}^{i^*}_R  \tilde{d}^j_R ,
\label{eq:leftleftnonuniv} \eeq
generally lead to
off-diagonal squark masses in the
quark mass eigenstate basis.
Left-right squark masses,
\beq  \sum_{i j} \d \tm^2_{i_L j_R} \tilde{d}^{i^*}_L  \tilde{d}^j_R +H.c .,
\label{eq:leftrightsquarkmasses} \eeq
are obtained from scalar trilinear couplings to
Higgs doublets.
In general, these also lead to off-diagonal squark masses
in the quark mass eigenstate basis\footnote{In general, the quark and
left-right squark mass matrices will not be proportional},
\beq \d \tm^2_{d_L s_R} \tilde{d}^*_L  \tilde{s}_R + \d \tm^2_{d_L
b_R} \tilde{d}^*_L  \tilde{b}_R +
\d \tm^2_{s_L b_R} \tilde{s}^*_L  \tilde{b}_R
+ L \longleftrightarrow R . \label{eq:
leftrightphyssquarkmasses} \eeq
The assumption of near degeneracy of down squark masses,
generally required by flavor-changing neutral current
(FCNC) constraints\footnote{Strictly
speaking, near degeneracy is required
among the left-
handed squarks and among the right-handed squarks separately.
The degeneracy
requirement can be
satisfied \refmark{dkl} or relaxed \refmark{nirseiberg}
in models with horizontal symmetries.} \refmark{nilles}
for $m_{\tq}$ and $m_{\tg}$ of a $TeV$ or less,
corresponds to $\d \tm^2 <<  m^2_{\tq}$.   This allows us to work in
the squark mass insertion
approximation when computing radiative flavour-changing effects.
We neglect CP violation and take
all
masses and operator coefficients to be real.

We will be interested in
contributions to the chromomagnetic dipole operators which
are generated
by the gluino penguin graphs of Fig. 5.\footnote{Neutralino
penguin contributions are suppressed by ${\cal O} ({\alpha_{em}
\over
\alpha_s})$.  Chargino and charged Higgs dipole penguin
contributions
to the \dI amplitude and $\B$ suppression must also be substantially
smaller in the MSSM
due to various factors, including small Yukawa couplings
and FCNC constraints.}
These graphs were first studied in ref. \refmark{gerard}
in the context of potential contributions to $\e' /  \e$
and have also been studied in the context of $b$ decays
\refmark{bertollini}.
The resulting chromomagnetic dipole
operator coefficients in the
quark interaction basis, see eqs. (\ref{eq:Lginteraction}),
(\ref{eq:Qginteraction}),
are given at ${\cal O }({\d \tm^2 / m_{\tq}^2 })$ by
\beq  C_{\scriptstyle G} ^{i j}  (m_{\tq})  =  {{\a_s   x }\over
{8 \pi
m_{\tg}}}\left(3E(x) -{16 \over 3}
C(x)\right)
 {\d \tm^2_{i_L j_R} \over m_{\tq}^2} ,
 \label{eq:susycginteraction} \eeq
where $x={m^2_{\tg} /  m^2_{\tq}} $, (at
$m_{\tq} $).
The loop integrals
$E(x)$ and $C(x)$, corresponding to vector
boson
emmission
from the gluino and squark lines,
respectively, are
given by
\bea    E(x) & = & {1 \over  (1-x)^3}
\left[2(1-x)
+(1+x){\rm ln}x \right]
 \\
  C(x) & = & { 1 \over  {4(1-x)^4}} \left[5x^2
-4x -1
-2x(x+2){\rm ln}x \right].
\eea
Associated radiative contributions
to the down quark
mass matrix in the quark interaction basis, see eq.
(\ref{eq:Lminteraction}), are given at
${\cal O }({\d \tm^2 / m_{\tq}^2 })$ by
\beq {\scriptstyle \Delta } m_{ij} (m_{\tq} ) = {4
\over 3} {\a_s \over  {2 \pi}} {\d \tm^2_{i_L j_R} \over m_{\tq}^2}
m_{\tg}{{(x {\rm ln} x +1-
x)}\over
{(1-x)^2}}. \label{eq:mfsusyinteraction} \eeq

Radiatively induced dipole operator coefficients and
quark masses in the physical quark basis
are given in terms of the corresponding left-right squark
mass matrix entries.
For example,
\bea  \cgsp (m_{\tq}) & = & {{\a_s   x }\over
{8 \pi
m_{\tg}}}\left(3E(x) -{16 \over 3}
C(x)\right)
 {\d \tm^2_{d_L s_R} \over m_{\tq}^2}
 \\
\dmdsp & = &
{4
\over 3} {\a_s \over  {2 \pi}} {\d \tm^2_{d_L
s_R} \over m_{\tq}^2} m_{\tg}{{(x {\rm ln}x +1-
x)}\over
{(1-x)^2}} .
\label{eq:susyamp} \eea
The loop integrals $E(x)$ and $C(x)$ correspond to gluon emmision
from the gluino and squark lines, respectively.
$\cgsm$ and $\dmdsm$ are obtained
via the substitution ${\d \tm^2_{d_L
s_R} \to {\d \tm^2_{d_R
s_L}}}$.
Expressions for the other chromomagnetic
dipole operator coefficients and quark masses in
eqs. (\ref{eq:lagrangian}) and (\ref{eq:masslagrangian})
are completely analogous.

Note that whereas our model-independent analysis was restricted
to the case of
a single exchange of particles in the loop,
up to six squark mass eigenstates can be
exchanged in the supersymmetric loops, leading to matrices
$C^{ij}_{\scriptstyle G}$ and
${\scriptstyle \Delta } m_{ij}$ which are generally rank 3.
Nevertheless, to good approximation these
two matrices are proportional,
given approximately degenerate squark masses.  Deviations from
proportionality first arise at
${\cal O }({\d \tm^4 / m_{\tq}^4 })$ and can be neglected for our
purposes.
The supersymmetric results can therefore be recast
interms of our model-independent parametrization, as in eq.
(\ref{eq:cijmij}).  In particular, $\zeta_{\scriptstyle G} $
is given in terms of ratios of loop
integrals and is flavor independent, depending only
on $m_{\tg}$ and $m_{\tq}$.   The scale of new physics, $M$,
is identified with the larger of the two masses.  As will become clear
below, maximization of the \dI amplitude favors $m_{\tq} >>
m_{\tg}$
so that $M$ will be identified with the squark mass scale.

\vglue 0.3cm
\noindent{ \bf  The \dI Amplitude.}
\vglue 0.2cm

We begin by estimating upper bounds on the
dipole induced \dI amplitude implied by
the observed mass difference, $\dmk^{exp} $.
The relevant supersymmetric contributions
\refmark{dugan,gabbiani,kelley} to
$\dmk$ are given in eq. (B1) of
Appendix B.
The matrix elements are evaluated in the vacuum
insertion approximation with $m_s$ and
$m_d$ taken
at $m_c$.  We choose $m_s =
150~MeV$ and
$m_d = 8~MeV$.
$\alpha_s$ and the
supersymmetric mass
parameters are taken
at the squark mass scale, $m_{\tq}$, and
QCD running of the $\Delta S=2 $ operator
coefficients to hadronic scales is not included.
These are clearly only order of magnitude estimates and
a more sophisticated treatment taking into account
QCD corrections and a more rigorous determination of the matrix
elements
is left for future work.

The first three terms in eq. (B1) depend on
the same squark mass insertions which enter the dipole operator
coefficients, $\cgs$.
Constraints on the chiral structure of the
\dI Lagrangian pointed out in the previous section require
$\abs{\d \tm^2_{d_L
s_R} } \gtap (2-4) \abs{\d \tm^2_{d_R s_L} } $ for large dipole-induced
contributions.
This suggests that the
$\d \tm_{d_L s_R }^4 $ term in eq. (B1) is the most important
for constraining the magnitude of the induced \dI amplitude.
The sign of its contribution
to $\dmk$ is
the same as the standard model
contribution
and the observed mass difference.
However, both the
$\d \tm_{d_L s_L }^2 \d \tm_{d_R s_R }^2$ and
$\d \tm_{d_L s_R }^2 \d \tm_{d_R s_L }^2 $ terms
can have opposite sign and can
compensate.\footnote{Note that the integrals
$f_6 (x) $ and $\tilde{f}_6 (x)$ have opposite sign,
while the various squark mass insertions
can either be positive or negative.}

Inorder to study the relative importance
of the first three terms in eq. (B1) we equate, separately, the
magnitudes of
the first and third
terms to
$\dmk ^{exp}$ and
plot the corresponding upper bounds on ${\d
\tm^4_{d_L s_R} /
m_{\tq}^6}$ and ${{\d \tm^2_{d_L s_R} \d \tm^2_{d_R s_L}} /
m_{\tq}^6}$, respectively, in Fig. 6 as a function of $x$.\footnote{Since
gluino and squark masses are not fixed by $x$ we take
$\alpha_s = .11 $ in obtaining Fig. 6, which is a reasonable
approximation for weak or $TeV$ scale squarks.}
For $ x \sim .01 $ to 1 (corresponding to weak scale gluinos and
weak to $TeV$ scale squarks) and
dominance of $\Qgsp $,
Fig. 6 confirms that the first term in eq. (B1)
provides the most important constraint on the induced \dI
amplitude.
Its contribution to $\dmk$
would be considerably larger than that of the next two terms.
However, for $x \ltap .01$ substantial cancelations are possible between
the first two terms and the third term.

In Fig. 7a we plot upper bounds from $\dmk$
on the
contribution of $\Qgsp$ to $R_0$,
the ratio of the induced \dI amplitude
to the observed amplitude, for weak scale gluino
masses.  The bounds correspond to settting the first term in
eq. (B1) to $\dmk^{exp}$, $2\dmk^{exp}$ and $3\dmk^{exp}$.
The more
liberal bounds take into account the possibility of
accidental cancelations, up to 1 part in 3 - 4, among the supersymmetric
contributions to
$\dmk$.
$\cgsp$ is evolved from $m_{\tq}$ to $m_c =1.4~GeV$
taking all relevant
thresholds, including $m_{\tg}$, into account.

According to Fig. 7a the induced \dI amplitude can account for
$30\% $ to $50 \%$ of the observed amplitude if the
unknown suppression factor ${m_K^2 / \Lambda ^2}$
lies in the range
.2 to .4, as suggested in the model-independent analysis.
This is especially true for lighter gluino masses
or for small
$x$ because the loop integral $E(x)$,
associated with the larger of
the two contributions in eq. (\ref{eq:susyamp}), gluon emmision from the
gluino line,
increases substantially as $x$ decreases.
Note that an accelerator lower limit on
the
gluino mass is difficult to obtain since
gluino
cascade decay depends on many parameters.
Although a strict lower limit is close to $95~GeV$,
it is more likely to be around $125~GeV$
\refmark{howie}.

In Fig. 7b we plot upper bounds on the mass parameter
\beq \tm_{{d_L}s_R} \equiv {\d
\tm^2_{{d_L} s_R}
\over
m_{\tq}}, \label{eq:tmds} \eeq
corresponding to the bounds
in Fig. 7a.
$\tm_{{d_L}s_R} $ essentially measures the
amount of $SU(2)_L$
breaking
contained in ${\d
\tm^2_{{d_L} s_R}}$.
It should
not be much
larger than the weak scale, based on the
requirement
that massive Higgs-squark scalar
trilinear coupling coefficients should be less
than or
of order the
squark mass scale in order to prevent $SU(3)_C$ breaking
\refmark{nilles}.
{}From Fig. 7b it follows that, for the gluino
masses
we've chosen, the squark
masses can not be much larger than 2 or
$3 ~TeV$ when saturating the $\dmk$ bounds.
According to Fig. 7a  this is not very restrictive as far as
\dI enhancement is concerned.
Note that Fig. 7b confirms the validity of the
squark mass
insertion approximation in the region of squark masses of
interest.

In Fig. 7c we
study implications of \dI enhancement for the
quark mass spectrum.
Upper bounds on the induced quark mass $\dmdsp (m_c) $,
correponding to the bounds of Fig. 7a, are plotted in order to
probe dependence on the gluino and squark masses.
According to Figs. 7a and 7c
generation of $\theta_c $ or $m_s$ (corresponding to
$\dmdsp \sim 35~MeV$)
together with a
large dipole-induced \dI amplitude
favors lighter gluino masses, $m_{\tg} \sim 125~GeV$
to $175~GeV$,
and lighter squark masses, $m_{\tq} \sim {1\over 2}~TeV$.  Note
that $m_{\tq}$ in this range corresponds to {\it{Region II}}
of our model-independent analysis.\footnote{Given
$\dmdsp \sim \theta_c m_s $, generation of $\theta_c $
would correspond to $ {\d \tm^2_{1_L 2_R}} \sim {\d\tm^2_{d_L
s_R}}$
while generation of $m_s $ would correspond to
${\d \tm^2_{2_L 2_R}} \sim {\d\tm^2_{d_L s_R} \over \theta_c }$.
According to Fig. 7b, ${\d \tm^2_{1_L 2_R} \over m_{\tq} }$ and
${\d \tm^2_{2_L 2_R} \over m_{\tq}}$ would be
sufficiently small in each case when compared to the weak scale.}

To make further contact with the model-independent
analysis we plot
contours of constant $R_0$ in the
($\vert{\xi_{ds}^+}\vert$,$m_{\tq} $) plane of Fig. 8.
Contours of constant $\dmk$, again corresponding to contributions of
the first term in eq. (B1), are also included in order to determine
the allowed regions of the plane. $m_{\tg} =150~GeV$ is chosen for illustrative
purposes,
reflecting the tendency towards larger \dI amplitudes at
lower gluino masses.
$\B$ contours are included for later comparison.

The similarities between Fig. 8 and Fig. 4
demonstrate that supersymmetry can provide a
realization of our
model-independent conclusions.
In particular, we see that in Region II
the induced \dI amplitude can
reasonably account for $75 {m_K^2 \over \Lambda ^2 }~ \% $
to
$150{m_K^2 \over \Lambda^2}  ~ \%$ of the observed amplitude,
in direct association with generation of  $\theta_c$ or $m_s$
($\xi_{ds}^+ \sim 1$).
However, the larger \dI amplitudes may require
a one part in 3 - 4 cancelation among
the supersymmetric contributions
to $\dmk$.
Again, we remind the reader that our estimates of the latter are
fairly crude, especially since the
vacuum saturation approximation has been used.
Alternatively, for larger squark masses \dI enhancement will require
a small tuning of $\theta_c$ or $m_s$.
Finally, we have not taken into
account the potential contribution of
$\Qgsm $ to the \dI amplitude.
As previously noted, $K \to 3 \pi $ constraints probably
allow ${\cal O}(10 \%) $
of the observed amplitude to come from this source.  Associated
contributions to $\dmk$ from the second and third terms in eq. (B1)
would be sufficiently small.
The amplitudes generated by $\Qgsm$ and $\Qgsp$
can, a priori, add constructively,
strenghthening our conclusion that the
chromomagnetic dipole moments
could account for $30\%$ to $50\%$ of
the observed \dI amplitude.

\vglue 0.2cm
{\noindent{\it{Ultra-light Gluinos}}}
\vglue 0.2cm

Finally, we consider \dI enhancement for
gluinos in the `light-gluino window'
\refmark{lightgluinoexp,A0A2exp,clavelli,kuhn}, corresponding
to $m_{\tg} \sim
1$ to $4~GeV$, or $x \ltap 10^{-3}$ for weak scale
or heavier squarks.
Here we are motivated by the observation that
the allowed dipole-induced \dI amplitude increases with
decreasing $x$. It has been claimed that light gluinos would
also lead to
better agreement between the LEP
measurement of
$\alpha_s (M_Z)$
and  determinations of $\alpha_s $ at lower
energies \refmark{clavelli,kuhn} since they
would
slow the running of
$\alpha_s$ below $M_Z$.  Whether there
really is a
discrepancy between the proper extraction of
$\alpha_s $ from LEP and other experiments, or
whether parts of the light-gluino window are
actually not ruled out
\refmark{lightgluinocontroversy}
are issues which have
become increasingly controversial of late
about which we have nothing further to add.

In Figs. 9a, 9b and 9c we plot
upper bounds on $R_{0}$,
$\tm_{{d_L} s_R}$ and $\dmdsp(m_c)$,
respectively,
obtained as usual from contributions of the first term
in eq.
(B1) to $\dmk$.
Following ref.
\refmark{kuhn}, we choose $\a_s
(M_Z) = .124$, and evolve downwards at two-loops taking into
account
all relevant thresholds.
We see that for squarks in the $200~GeV$ to $400~GeV$
range, $R_0 \sim  {m_K^2  \over \Lambda^2}$ to
$2 {m_K^2 \over \Lambda^2} $ can be obtained
with little or no tuning
of $\dmk$, whereas $R_0 \sim 3 { m_K^2 \over  \Lambda^2}$
may require a moderate one part in three to four tuning.
So for ultra-light gluinos, $\Qgsp$ could
account for at least half of the observed \dI amplitude.
However, according to Fig. 6, a hierarchy of ${\cal O}(30)$ between
${\d \tm^2_{d_L s_R} }$
and ${\d
\tm^2_{d_R s_L} }$ would be required in order to satisfy
$\dmk$ constraints.  This condition
is discussed further in Appendix A.

The radiatively
induced quark masses are generally too small
to be relevant, with
$\dmdsp \sim 1~MeV$ to
$10~MeV$ typical.  A possible exception
arises for squarks near the $TeV$
scale.
For example, $\dmdsp (m_c) \sim \theta_c m_s $ can be obtained
for
$m_{\tq} \sim 800~GeV$ and $m_{\tg} \sim 4~GeV$.
Unfortunately, according to Fig. 6, a very large hierarchy of ${\cal O} (300)$
would
be
required between ${\d \tm^2_{d_L s_R} }$
and ${\d
\tm^2_{d_R s_L} }$, and
$\tm_{{d_L} s_R}$ would have to lie in the $200~GeV$
to $300~GeV$ range, which is on the high end
for an $SU(2)_L $ breaking squark mass.
So although ultra-light gluinos are promising for \dI
enhancement, this case does not conform to the conclusions of
our model-independent
analysis regarding quark
mass generation.  Because of the extreme values of $x$,
$\zeta_{\scriptstyle G}$ is substantially larger than 1,
contrary to what is naively expected,
so that large \dI enhancement is associated with relatively small
induced quark masses.

\vglue 0.3cm
\noindent{ \bf   Suppression of $\B$ and Radiative $B$ Decays.}
\vglue 0.2cm

Next we discuss supersymmetric
generation of the chromomagnetic dipole
operators $\Qgb$ and
$\Qgd$
via the $b$ penguin analogs of Fig. 5.  These
diagrams have been studied extensively in the past
\refmark{bertollini}.  We will see that they
can
resolve the discrepancy
between the measured value of $\B$ and the
parton model prediction in the standard model and that
this has
rich implications for the quark mass
spectrum and
radiative $B$ decays.  Again we will consider both
weak scale and ultra-light gluinos.

Expressions for
the operator coefficients $\cgb$, $\cgd$, and for
the radiatively induced masses
$\dmsb $, $\dmdb$ follow by analogy from eq.
(\ref{eq:susyamp}).
$\Gamma (b \to sg )$ and
$\Gamma (b \to dg)$ follow from  eq.
(\ref{eq:hadronicdecay1}).
The electromagnetic dipole operator
coefficients are
given by
\bea  \cfbp (m_{\tq}) & = & -{{2\a_s }  \over
{3
\pi}} {x \over m_{\tg} }  C(x)
 {\d \tm^2_{s_L b_R} \over m_{\tq}^2}
  \\
 \cfbm (m_{\tq}) & = & -{{2\a_s }  \over {3
\pi}} {x
\over m_{\tg} }  C(x)
 {\d \tm^{2 }_{s_R b_L} \over m_{\tq}^2} ,
\label{eq:cfbsusy} \eea
and similarly for the coefficients $\cfdp$
and
$\cfdm$.  Expressions for
$\Gamma (b \to s\gamma )$
and
$\Gamma (b \to d\gamma)$ follow from eq.
(\ref{eq:photonicdecay}).
Finally,
the supersymmetric box graph contributions to
$\Delta m_{\scriptstyle B} \equiv
m (B_d^0) - m (\overline {B}_d^0 )$ are given
in eq. (B2).

We are interested in suppression
of $\B$ due to
the inclusive gluon channel decay width, as
in the model-independent plots of
Figs. 2 and 4.
In Fig.10a we plot contours of constant $\B$,
for weak scale gluino masses,
in the plane of ${\scriptstyle \Delta} m' (m_c)$ vs. $m_{\tq}$
(${\scriptstyle \Delta} m' $
was defined in (\ref{eq:dmell})).
Although ${\scriptstyle \Delta} m' $
is proportional to $V_{cb}$
along these contours, this dependence and the accompanying
uncertainty
drop out for $\xi'$ (defined in (\ref{eq:xiell})).  In Fig. 8
contours of $\B$ have been included in the ($\xi'$,$m_{\tq}$) plane
for $m_{\tg} = 150~GeV$.
According to Fig. 10a or Fig. 8, supersymmetry can provide a
realization of the
model-independent conclusions of Fig. 2:
\vglue 0.2cm
\noindent{(i)} In Region I, corresponding to $m_{\tq} \sim 1~TeV$ to $2 ~TeV$,
the desired $\B$ suppression is associated with
$\xi' \sim 1$ for a wide range of gluino masses.
\vglue 0.2cm
\noindent{(ii)} In Region II, corresponding to $m_{\tq} \sim 300~GeV$ to
$700~GeV$, $\B$ suppression is associated with $\xi' \sim .1$ for gluino masses
below
$200~GeV$.
\vglue 0.2cm
\noindent{Potential} implications for the quark mass spectrum have
been discussed in Section 3 and Appendix A.  Restrictions special to
the
supersymmetric case are discussed below.

To check that the $SU(2)_L$
breaking squark mass insertions responsible for
$\B$ suppression are not too large
we define the mass parameter
\beq \tm' \equiv {{(\delta \tm_{s_L b_R} ^2
+ \d\tm_{b_L s_R} ^2+ \d\tm_{d_L b_R} ^2 +
\d\tm_{b_L
d_R} ^2 )} \over m_{\tilde q} },\label{eq:tmell}\eeq
and plot contours of constant $\B$
in the ($\tm' $,$m_{\tq}$) plane of Fig. 10b.
We also include upper bounds on $\tm_{d_L
b_R} $, obtained by setting the analog of the first term in
eq. (B1) equal to $m(B_d^0) - m(\overline {B}_d^0 )$.
Comparison of Figs. 10a and 10b confirms that
$B_d^0 - \overline {B}_d^0 $ mixing does not
significantly constrain $\B$ suppression.
However,  limitations on the size
of $SU(2)_L$ breaking squark mass insertions determine which
features of the quark mass spectrum can be accounted for.
In particular, in Region I, $\xi' \sim 1$ can be associated with
generation of $V_{cb}$ but not with the alternative, generation of
$m_b$.  In Region II, $\xi' \sim .1$ can be associated with
generation of $V_{ub}$ but not with the alternative,
simultaneous generation of $V_{cb}$ and $m_b$.\footnote{According
to
Fig. 10b, in the ruled out scenarios
${\d \tm^2_{3_L 3_R} / m_{\tq} } $ would be much
larger than the weak scale.}

In Region II large hierarchies are not required among the left-right down
squark mass insertions\footnote{
$\d \tm^2_{i_L j_R} \over m_{\tq} $
can be of order a few GeV for all i,j.}.
Equivalently, all entries of the radiatively induced quark mass matrix,
${\scriptstyle \Delta }m_{ij}$, can be of order $\theta_c m_s$
or $V_{ub} m_b$, therebye
accounting for $\theta_c$, $V_{ub}$, and $m_d$.

To study implications for radiative $B$ decays,
contours of constant $\B$
are drawn in
the
plane of ${\rm BR} (b \to  x \gamma )$ vs. $ m_{\tq} $.
These contours are essentially independent of $V_{cb}$.  Only the
supersymmetric contributions to $b \to x \gamma $ are taken into
account  but,
a priori, the standard model contributions could add constructively
or destructively.
According to Fig. 10c, the following can be concluded:
\vglue 0.2cm

\noindent{$\bullet$} In Region I, with $\xi' \sim 1$, new contributions to
$\BF$ tend to lie below
the standard model contribution, $(2-3) \times 10^{-4}$,
unless the gluinos are heavy.  Eq. (\ref{eq:xione}) implies that
contributions to
$\BFd$ will be two orders of magnitude smaller, as
in the standard model.
\vglue 0.2cm
\noindent{$\bullet$} In Region II, with $\xi' \sim .1$,
new contributions to ${\rm BR} (b \to x \gamma)$
are ${\cal O} (10^{-4})$.   If ${\scriptstyle \D} m_{db}^+ \sim
V_{ub} m_b $, as in eq. (\ref{eq:xitenth}), and as suggested
by the quark mass spectrum, then
$\BFd \sim (.1-1) \times 10^{-4} $,
a dramatic departure from the standard model.

\vglue 0.3cm
{\noindent{\it{ Ultra-light Gluinos}}}
\vglue 0.2cm

As in the case of \dI enhancement we end discussion
of $\B$ suppression with the case of gluinos in the `light-gluino'
window.  Figs. 11a,b are the analogs of Figs.
10a,b, respectively,
for $m_{\tg} = 1~GeV$ and $4~GeV$.
$\B \sim 10\% -11\% $ is readily
obtained, and $B_d^0 - \overline {B}_d^0 $
mixing and $SU(2)_L $ breaking constraints on the squark mass
insertions are easily satisfied.
However, as for \dI enhancement with light gluinos,
the quark mass
contributions do not play a significant role except
perhaps for squark masses in the $800~GeV$ to $1~TeV$ region.
In this case $\dmdbp \sim 20~MeV$, or ${\cal O} (V_{ub} m_b )$,
could help account for $V_{ub}$.  Finally, new contributions to
${\rm BR }(b \to x \gamma)$ depend weakly on $m_{\tq}$
and are ${\cal O} ( 2-3 ) \times 10^{-
5}$.  All left-right squark mass insertions can be of same order
since the radiative quark mass contributions are small,
implying that $\BFd$ can be an order of magnitude larger than in the standard
model.

\vglue 0.6cm

To summarize, comparison of
Figs. 4 and 8 reveals that {\it{Region I}} or {\it{Region II}}
dipole-operator
phenomenology can be realized in supersymmetric models with weak scale
gluinos. In particular, it is possible to tie in
$\B$ suppression with radiative generation of $V_{ub}$ (Region II)
or $V_{cb}$ (Region I). It should also be possible to tie in $30\%$ - $50\%$
of the \dI amplitude with radiative generation of $\theta_c$ or
$m_s$ (Region II).  For larger squark masses the $\dmk$ constraints are
weaker, but a small tuning of $\theta_c$ or $m_s$ would be required.
For ultra-light gluinos, $\B$ suppression and larger \dI amplitudes
are possible, but it is difficult to relate these effects to the quark
mass spectrum.
Finally, $\B$ suppression in Region II is associated
with large contributions to $\BFd$,
lying in the range $(.1 -1) \times 10^{-4}$.  For ultra-light gluinos,
contributions of order $(1-3) \times 10^{-5}$
are possible.

Supersymmetric models of chromomagnetic dipole operator phenomenology face
difficulties in
supergravity theories with general K\"{a}hler potential, or in
string theories with moduli-driven supersymmetry
breaking \refmark{louis}.\footnote{Supergravity
theories with minimal K\"{a}hler potential,
or string theory scenarios with dilaton driven \refmark{louis,moretti}
supersymmetry breaking can not generate large
enough dipole operator coefficients.}
In particular, the expected magnitudes \refmark{kagansugra,dkl} of
off-diagonal left-right squark
mass insertions will be too small to accomodate Region I
phenomenology, and will rule out any
chance for a connection to the quark spectrum with ultra-light gluinos.
Those mass insertions which involve the 3rd generation, $\delta
\tilde{m}^2_{1_L
3_R}$,
$\delta \tilde{m}^2_{2_L 3_R}$, etc., are expected to be ${\cal O}(m_{susy}
m_b)$,
which is large enough to obtain $\B$ suppression and $V_{ub}$
in Region II, or $\B$ suppression with ultra-light
gluinos.
Finally, the left-right squark mass insertions of relevance
to the \dI amplitude are expected to be ${\cal O}(m_{susy} m_s)$
which is about an order of magnitude smaller than required for
significant enhancement with weak-scale or
ultra-light gluinos.
However,  these estimates of the mass insertions are uncertain
by at least a factor of 3, since there are many dimensionless parameters in
the Kahler potential which could be ${\cal O} (1)$.
Larger contributions to these mass insertions may also
arise if `hidden sector', or
string-moduli fields couple to the observable sector
via non-renormalizable terms.\refmark{weldon}

It is suggestive that in supersymmetric models
radiative mass contributions associated with
$\B$ suppression or \dI enhancement are often of the right
magnitude
to account for several features of the quark mass spectrum.
However,
it remains to construct supersymmetric models in which they
provide a unique origin for these features.  In particular,
one
would have to show that supersymmetry breaking can lead to large
enough flavor symmetry breaking in the
squark sector in models in which tree-level Higgs Yukawa couplings
are not important for the light quark spectrum.

In contrast, the dipole operators are often a necessary
outcome of quark mass generation in technicolor models, or in models of
quark and lepton substructure.
Next, we discuss a class of technicolor models with large
chromomagnetic dipole moments.

\vglue 0.6cm
{\elevenbf\noindent 5.  Techniscalar Models}
\vglue 0.2cm

We begin with a brief description of
techniscalar models \refmark{eotvos}.
Unlike in
extended technicolor (ETC)  models \refmark{etc},
the technicolor gauge group is not extended
to a horizontal group.  Instead, the ETC gauge bosons are replaced with
technicolored scalars (techniscalars).
Flavor-changing
neutral currents first arise at the one-loop
level and are suppressed.
Furthermore, the quark-techniscalar-
technifermion Yukawa couplings can vary substantially.  These
features allow
the masses of all techniscalars to
be of order $1~TeV$.  In contrast, ${\cal O} (100~TeV)$ masses
are required for ETC gauge bosons
which couple to the light quarks.

Consider the gauge group $G= SU(N)_{TC} \times
SU(3)_C
\times
SU(2)_L \times U(1)_Y$, together with
three quark families and the following technicolored fields:
a right-handed $SU(2)_L $
doublet of technifermions
$T_R (N,1,2,0)=(U_R , D_R )^T $, two left-handed $SU(2)_L$ singlet
technifermions
$U_L (N, 1,1,1/2)$,
$D_L (N, 1,1,-1/2)$, all with charges $\pm {1\over 2}$, and a charge
$ {1 \over 6 }$ techniscalar $\omega
(\overline{N}, 3,1,1/6)$.  Transformation properties with respect to
the technicolor group, $SU(N)_{TC}$, and the standard model
gauge group have been included in parenthesis.  The most general quark Yukawa
couplings are given by
\beq  {\cal L}_{Y} = {\bf{h}}_i
 \om    {\overline {  Q^i_ L }} { T_ R }
+{{\bf{ h}}^u_i}^{\dagger} \om ^* {\overline{ U_L}}{ u^i_{R}}
+ {{\bf{h}}^d}^{\dagger}_i\om ^* {\overline{D_L}}{ d^i_{R }}  +H.c.,
\label{eq:tcyukawas} \eeq
where $\bf{h}^u$, $\bf{h}^d$ and $\bf{h}$ are
dimensionless three-component Yukawa coupling
vectors and
latin indices label the quark interaction basis states.
$\om$ acquires an explicit mass from the scalar sector of the Lagrangian
and a `constituent' mass from technicolor dynamics.\footnote{
Scalar technicolor models can be supersymmetrized
in order to protect the masses of the scalars.  In turn, supersymmetric
FCNC can be
suppressed since a multi-TeV supersymmetry breaking scale is natural in this
framework \refmark{dks}. }
As usual we ignore CP violation and take all parameters to be real.

Technifermion condensates will induce up and down
quark mass matrices via techniscalar exchange.
In the limit $m_{\om} >> \Lambda_{TC}$, where $\Lambda_{TC} \sim 1~TeV$,
the down mass matrix, in the interaction basis, is
given by
\beq {\scriptstyle \Delta} m_{ij}  \approx
{\bf{h}_i}{{{\bf{h}}^d
}^{\dagger}_j} {{\langle \overline {D} D \rangle } \over {4m_{\om}^2
}}.\label{eq:tcmasses} \eeq
The up quark matrix is analogous.
The technifermion condensates are
estimated to be \refmark{manohar}
\beq \langle \overline {D} D
\rangle = \langle \overline {U} U
\rangle
\sim \left({3\over N_{TC}}\right)^{1\over 2} 4 \pi
\left({v \over \sqrt{N_D}}\right)^3~GeV^3 ,
\label{eq:condensate} \eeq
where
$v=246~GeV$, and $N_D$ (equal to 1 above) is the number of technifermion
doublets.

Chromomagnetic dipole operators are due to emmision of a gluon
by the exchanged techniscalar.  The down quark
coefficients are given by
\beq C_{\scriptstyle G}^{ij}   \approx
{{{\scriptstyle \Delta}
m_{ij} } \over
{2 m_{\om}^2}}  \label{eq:tccoefficients} \eeq
at $m_{\om}$.
Note that $C_{\scriptstyle G}$ and ${\scriptstyle \Delta} m$
are proportional, rank 1 matrices.
In terms of the general parametrization in (\ref{eq:cijmij}), the above example
corresponds to
$\zeta_{\scriptstyle G} \approx {1\over 2}$.
To estimate the physical (flavor-changing) chromomagnetic dipole moments
we insert this value into the model-independent expressions in
(26),(27) for
$\cgds$, $\cgsb$, and $\cgdb$, and identify the scale of new
physics $M$ with $m_{\om}$. The renormalization scale factor $\eta (\mu)$
is the same as in eq. (\ref{eq:cijmij}) for non-supersymmetric models.

We will mainly be interested in the case $m_{\om} \sim \Lambda_{TC}$,
especially for $m_{\om}$ in
Region II.
Unfortunately, it is difficult to calculate the induced quark masses and
dipole moments in this case since strong technicolor dynamics become
important.  For example, it may be that
the exchanged techniscalar and technifermion
bind so that the quark's mass can be interpreted as due to mixing
with a composite heavy quark.
Nevertheless, we expect the above expressions to give the correct orders of
magnitude, and we still expect proportionality between
${\scriptstyle \Delta} m$
and $C_{\scriptstyle G}$.

In models with a minimal set of technifermions and a single techniscalar
only a single family acquires masses from techniscalar exchange,
which we identify with $m_t$ and $m_b$.\footnote{One obtains
$m_t \approx \abs{\bf{h}} \abs{\bf{h}^u}
\langle {\overline U} U \rangle /4m_{\om}^2 $,and $m_b \approx \abs{\bf{h}}
\abs{\bf{h}^d}
\langle {\overline D}D \rangle /4m_{\om}^2 $. }
Note that $SU(2)_L$ invariance automatically aligns the left-handed top and
bottom mass eigenstates (at
0'th order in
light quark masses), as required by the KM hierarchy.
Exchange of an additional techniscalar (copy of $\om$)
or an additional set of technifermions (copies of $T_R$, $U_L$,
$D_L$ ) with smaller Yukawa couplings can generate a second unit-rank matrix
for
up quarks and for down quarks, with eigenvalues of order $m_c$ and $m_s$,
respectively.  This time
$SU(2)_L$ invariance insures approximate alignment of the left-handed
charm and strange eigenstates, as required
by $\theta_c$.  (Some details of these models are discussed in Appendix A.)
First generation masses, $\theta_c$ and $V_{ub}$
could also be due to techniscalar exchange or they could have an entirely
different origin.  For example,
additional supersymmetric or multi-scalar flavor physics could generate
radiative mass contributions of
${\cal O} (\theta_c m_s)$, as we saw in the previous section.
Alternatively, Higgs doublets can obtain
small vacuum expectation values by
coupling to the technifermions\refmark{tHooft,dks}.
If they are very heavy and, or, their couplings are small they
could simultaneously account for the light
quark spectrum and evade FCNC bounds.\footnote{
However, this option does not have a built-in mechanism for alignment of charm
and strange eigenstates.}

We begin with a phenomenological analysis of
\dI enhancement and $\B$ suppression due to exchange of
a single techniscalar-technifermion pair, leading to the masses and
chromomagnetic operator coefficients of eqs. (\ref{eq:tcmasses}),
(\ref{eq:tccoefficients}).
{}From this analysis we hope to learn which features of the quark mass
spectrum,
or which scenarios
outlined above, are most naturally associated with either phenomenon.

\vglue 0.4cm
\noindent{ \bf   \dI enhancement, $\B$, and radiative $B$ decays}
\vglue 0.3cm

In Fig. 12a contours of constant \dI enhancement, or $R_0$, in the
plane of $\abs{\xi_{ds}^+ - \xi_{ds}^- } $ vs. $m_\om $ are obtained
by taking $\zeta_{G} \approx {1 \over 2} $ in the
analysis leading to Fig. 4.
It should come as no surprise that
$\dmdsp \sim 30 - 50 ~MeV$, or  $\xi_{ds}^+ \sim 1$, singles
out {\it{Region II}} for substantial \dI enhancement, corresponding to
$R_0 \sim (1-1.5){m_K^2 \over \Lambda^2 }$.
An important question is whether technicolor dynamics allow $m_{\om}$ in Region
II?
Estimates of the technifermion constituent mass \refmark{chivukula}
(obtained by scaling the QCD constituent mass) give
\beq m_{TC} \sim (300~MeV) {v \over {\sqrt{N_D} f_\pi }},
\label{eq:mTC}
\eeq
or $800~GeV$ for $N_D =1$, $550~GeV$ for $N_D =2$.
Assuming that techniscalar `constituent' masses are of same order
we take as a reasonable bound $m_{\om} \gtap {1 \over 2} ~TeV$ which
of course would allow for solutions in Region II.

To examine the relevance, or lack thereof, of $\dmk$ constraints
it is necessary to define the Yukawa couplings of the left-handed and
right-handed quark mass
eigenstates to $\om$:
\beq
\lambda_q   \equiv \langle{ q_L}\vert {\bf h}\rangle,~~~\overline{\lambda}_q
\equiv
\langle {\bf h}^d \vert {q_R}\rangle,~~~~q=d,s,b.  \label{eq:tclambdas}
\eeq
The quark masses $\dmdsp$ and $\dmdsm$ are proportional to
$\lambda_d  \overline{\l}_s $ and ${\l_s}^* \overline{\lambda}_d^*$,
respectively.
To get rough estimates of new box-graph contributions to $\dmk$
in terms of these couplings
we ignore technicolor interactions and assign the technifermions in the loops a
mass equal
to
the technifermion `constituent mass', $m_{TC}$, in eq. (\ref{eq:mTC}).
The resulting expressions are given in eq.
(B2) of Appendix B.
In Fig. 12b we plot contours of $R_0$ in the
plane of $|\lambda_d \overline{\lambda}_s -
\overline{\lambda}_d \lambda_s |$ vs. $m_{\om} $, for $N_{TC} =4$ and
$N_D =1$.
We also include an `upper-bound' on
$\lambda_d \overline{\lambda}_s $,
obtained by setting the first term in eq. (B2)
equal to $\dmk^{exp} $.
Remarkably, and in spite of the crude nature of our estimates,
this result clearly indicates that $\dmk$ constraints are not
a factor in limiting \dI enhancement in techniscalar models, unlike
what we saw in supersymmetric
models.\footnote{$\dmk$ constraints for $N_D =2$ are slightly more restrictive,
but our conclusions would not change qualitatively.}
The remaining terms in eq. (B2)
are also not restrictive.  In fact, the main
constraint on \dI enhancement comes from $\theta_c$ or $m_s$.

For $\B$ suppression we are, as usual, interested in the
cumulative effects of $\Qgs$ and $\Qgd$.
Contours of $\B = 10\%, 11\%$ in the ($\abs{\xi'}$,$~m_{\om}$) plane can be
read
off
directly from the model-independent
plots of Figs. 2 or 4 by taking $\zeta_{G} \approx {1 \over 2} $,
and are included in Fig. 12a.
As expected, $\xi' \sim 1$ singles out Region I ($m_\om \sim 1-1.5~TeV$)
for $\B$ suppression, while
$\xi' \sim .1$ singles out Region II ($m_\om \sim {1 \over 2}~TeV$).

What about radiative $b$ decays in these models?
The electromagnetic dipole operators $\Qfb$
and
$\Qfd$ are induced by radiation of a photon
from
the exchanged techniscalar.
Similar contributions have been
considered
in extended technicolor models \refmark{Lisa}.
The operator coefficients $\cfb$ and $\cfd$
are given at $m_{\om} $ by
\beq \cfb \approx {Q_{\om} \over Q_d}
{\dmsb \over
2m_{\om}^2}, ~~~~ \cfd \approx {Q_{\om} \over Q_d}
{\dmdb \over
2m_{\om}^2}. \label{eq:tcbslcoefficients}
\eeq
Comparison with the model-independent parametrization
of Section 3, eqs. (\ref{eq:parametrization}), (\ref{eq:parametrizationF}),
gives ${\zeta_F \over \zeta_G} = {Q_{\om} \over Q_d}=-{1 \over 2}$.
According to Fig. 3, new contributions to
${\rm BR} (b \to x \gamma)$ associated with $\B$ suppression must
therefore lie below the CLEO bound.
This is borne out in Fig. 12c.
In both Region I and Region II one obtains
${\rm BR} (b \to x \gamma) \sim (2- 3) \times 10^{-4}$.
Given $\dmdb \sim
V_{ub} m_b $, we expect $\BFd \sim 10^{-2}\cdot \BF$ in Region I,
but in Region II, as usual, $\BFd$ is likely to
exceed the standard model prediction by an order of magnitude or more.
Finally, we mention that new contributions
to $B^o - \overline{B^o}$ mixing are too small to constrain
$\B$ suppression, as one would expect from the weakness of $\dmk$
constraints.

\vglue 0.3cm
\noindent{ \bf   Implications of Quark Mass Generation}
\vglue 0.2cm

Implications of quark mass generation for \dI enhancement and $\B$ suppression
are summarised below. Details are provided in Appendix A.
We assume that
$m_t$ and $m_b$ are due to techniscalar exchange, and that an
interaction basis exists in which the full down quark mass matrix respects
the hierarchy of eq. (A1):
\vglue 0.2cm
\noindent{\bf(a)}  If only 3rd generation masses are due to
techniscalar exchange (minimal
techniscalar model)
we expect $\dmsb \sim V_{cb} m_b$ and $\dmdb \sim V_{ub} m_b $, or
$\xi' \sim 1$.  This means that
$\B$ suppression must take place
in Region I. However, $\dmds$ would be too small to obtain substantial \dI
enhancement.
\vglue 0.2cm

\noindent{\bf(b)}  If a second set of technifermions is introduced
($N_D =2$)
in order to generate $m_c$, $m_s$ and
$V_{cb}$
then \dI enhancement and $\B$ suppression are naturally accomodated in
{\it{Region II}}.
The reason is that $C_G$ and ${\scriptstyle \Delta} m$ remain proportional
even though they are now of rank 2.  This means that $\dmsbp$ and $\dmsbm$
will be ${\cal O}(V_{ub}m_b)$ or ${\cal O}(\theta_c m_s)$, since they will be
determined by mass
contributions responsible for first
generation masses and mixing angles.
$\dmds$ and $\dmdb$ will also be of this order, implying that $\xi' \sim .1$
and
$\xi_{ds}^{\pm} \sim 1$, so
Region II phenomenology is singled out.  In particular, $30\% - 50 \%$
of the \dI amplitude may be attributable to $\Qgs$.
\footnote{As an illustrative example, consider ${m_K^2 \over \Lambda^2}
\approx .2$
and $m_{\om} \approx  {1 \over 2}~TeV$. According to Fig. 12a,
$\xi_{ds}^+ \sim 1$ and
$\xi_{ds}^- \sim -.5$ would generate $30 \%$ of the \dI amplitude.
Only $10 \%$ would be due to $\Qgdsm$,
so constraints on the chiral structure of the
\dI Lagrangian would be
satisfied.  If we further allow $\xi_{ds}^+ \approx 1.5$
(which requires less than a 1 part in 2 tuning between
different contributions to $m_s$)
, or ${m_K^2 / \Lambda^2} \approx .3$, then $50 \%$ of the \dI
amplitude would be accounted for.}
In general, there will be deviations from proportionality
of $C_G$ and ${\scriptstyle \Delta} m$ due to non-universal
technifermion current
mass\footnote{Technifermion current masses insure sufficiently heavy
technipions.
One possibilty for their origin is exchange of heavy gauge-singlet scalars,
leading to effective four-technifermion operators.} corrections
or radiative Yukawa coupling corrections to the technifermion propagators.
However, the propagators are dominated by large
and universal
constituent masses, so these corrections are generally too small
to alter our conclusions significantly.\footnote{
One might expect corrections to $\cgsb$ are ${\cal O} ({\delta m
\over m_{TC}}{ {m_b V_{cb}} \over m_{\om}^2 })$, where $m_{TC} \sim m_{\om}
\sim
{1 \over 2}~TeV$ and $\delta m $ is a typical technifermion current mass.
For example, $\delta m \sim 10 -50~GeV$
would produce ${\cal O} \left(
{200 \over \sqrt{N_D}}-{450 \over \sqrt{N_D}} ~GeV \right)$
technipion masses, but corrections to $\cgsb$
would be $\ltap {(20~MeV )
\over m_{\om}^2 }$ for $N_D =2$.}

\vglue 0.2cm

\noindent{\bf(c)} If 2nd generation masses and $V_{cb}$
are due to a second techniscalar,
$\B$ suppression again singles out Region I, as in (a).  Only in the limit of
degenerate or nearly degenerate techniscalar masses are \dI enhancement
and $\B$ suppression possible in Region
II, since $C_G$ and
${\scriptstyle \Delta} m$ would be nearly proportional, as in (b).
However, in the absence of additional (horizontal) symmetries
an order of magnitude tuning of techniscalar masses would be required.
\vglue 0.2cm

\noindent{\bf(d)} Finally, if 1st generation masses and mixing angles
are due to exchange of a second
(if there are already two sets of technifermions
responsible for 2nd and 3rd generation masses)
or third techniscalar, then they can be associated with
Region II phenomenology.\footnote{However, if both 1st and 2nd generation
masses
are due to exchange of the same techniscalar, but different
technifermions, $\cgdsp$ is suppressed and large \dI amplitudes are not
possible.}
If first generation masses and mixing
angles are not due to techniscalar interactions they might still
be associated with contributions to Region II phenomenology,
as we saw in the previous two sections.

\vglue 0.2cm

In techniscalar models the connection between the quark spectrum and
dipole operator phenomenology is transparent, and the potentially rich
phenomenological implications
of new flavor
physics, particularly in Region II, are well-illustrated.
Yet a further
consequence of a light ${1\over 2}~TeV$ techniscalar would be a large top
chromomagnetic
dipole moment, leading to
substantial enhancement of the Tevatron $t \bar t$ production
cross-section \refmark{rizzo}.  This is relevant in light of
recent evidence for top production at CDF \refmark{cdf}.

It stands to reason that important operators of different
dimension then the dipole moment operators may be generated
by new flavor physics.  We end this section with discussion
of dimension-6 operators
which impact on the decays $Z \to b \bar b $ and $b \to s \mu^+ \mu^- $.
In particular, we remark on the effects of the following
interaction between quark and technifermion $SU(2)_L$ doublets,
\beq {\Delta  {\cal L}}_6 \approx -{{{\bf{h}}_i
{\bf{h}}^{\dagger}_j }\over 4m_\om^2 }
({\overline {  Q_ L }}_i \gamma^\mu   {\tau}^a
{Q_L}_j )(\overline{T_R} \gamma_{\mu} \tau ^a T_R) ,
\label{eq:L6}\eeq
which is induced by techniscalar exchange.
The effects of similar dimension-6 operators have been considered
in ETC models \refmark{selipsky,Lisa,gates,soares,terning}.

For simplicity, we assume a minimal techniscalar
scenario, as in (a) above.
Replacing the technifermion current in (\ref{eq:L6}) by a sigma-model
current, as in \refmark{selipsky,georgitext},
and assuming the hierarchy in (A1), one  obtains the following
bottom quark couplings to the $Z$,
\beq  \Delta {\cal L}_Z \approx  {m_t \over 8 \pi v}
\abs{\lambda_t \over \overline{\lambda}_t } {e \over {sin \theta
cos \theta }}
(\overline {b_L} \gamma^\mu b_L  + {\cal O}(V_{cb}  ) \overline {s_L}
\gamma^\mu
b_L + {\cal
O}(V_{ub}  ) \overline {d_L} \gamma^\mu b_L) Z_\mu  .\label{eq:LZ}\eeq
Note that the $Zb_L b_L$ and $Z t_L t_L $ couplings increase in
magnitude\footnote{By
an amount $
{m_t \over 8\pi v}
\abs{\lambda_t \over \overline{\lambda}_t } {e \over {s_\theta c_\theta}}$.},
opposite to what happens in ordinary ETC models \refmark{selipsky}.
This is because the
technifermion current in
eq. (\ref{eq:L6}) is right-handed, giving opposite sign axial-vector couplings.
This is also the case in modified ETC models in which ETC gauge bosons
carry $SU(2)$ charge \refmark{terning}.

The resulting increase in $R_b \equiv {\Gamma_b
/ \Gamma_h }$, for $m_t \approx 170~GeV$, relative to
the standard
model prediction of $\approx .216$,
is estimated to be
\beq {\delta R_b \over R_b} \approx 9.9\% \abs{\lambda_t \over
\overline{\lambda}_t}
\left[ {m_t \over
170~GeV} \right]  . \eeq
The
LEP full fit \refmark{LEPRb}, $R_b = 0.2202 \pm .0020$,
corresponds to ${\delta R_b
\over R_b} <
2.9 \% (1\sigma ),$ $3.8\% (2 \sigma) $, so we require
${\lambda_t \over \overline{\lambda}_t} \sim {1\over
3}$.
Adapting the ETC analysis of $b \to s \mu^+ \mu^- $ in ref. \refmark{soares} to
the
techniscalar
case, we find
\beq {\rm BR} (b \to s \mu^+ \mu^-) \approx  9 \times 10^{-5}
\left({\lambda_t \over \overline{\lambda}_t} \right)^2 ,\eeq
or $\sim 10^{-5}$. This should be
compared to the present upper bound \refmark{mumubound} of $5\times
10^{-5} $, and the
standard model prediction \refmark{savage} of $\approx 6\times 10^{-6}$ for
$m_t
\approx
170~GeV$.
It is important to note that, quite generally, new contributions to $R_b$
and $b \to s \mu^+ \mu^- $ will be correlated as above.

\vglue 0.6cm
{\elevenbf\noindent 6. Conclusion
}
\vglue 0.2cm

We begin with a brief discussion of some of the relevant issues
which have not been addressed in this paper.  The first concerns CP violation
in
the Kaon
system.
Although we have set all CP violating phases to zero,
chromomagnetic
dipole operators can, in general, make
substantial contributions to ${\epsilon' / \epsilon}$.
In particular,  if they account for $30\% -50\%$ of the
\dI amplitude then the phases entering the dipole operator coefficients
must be extremely small, satisfying
\beq \abs{Arg [\cgdsp -\cgdsm]} \le (3-5) \times 10^{-4}
\abs{ {\epsilon' /\epsilon } \over 2 \times 10^{-4}}. \label{eq:CPphases}
\eeq
On the other hand, the measured value of $\abs{\epsilon}$ requires
$\abs{Arg[\dmk] } \approx 6 \times 10^{-3} $.
This hierarchy of phases poses a challenge for model-building efforts since
it suggests that flavor physics responsible for large dipole
operator coefficients can not be the source of CP violation in the Kaon
system, especially $\epsilon$.

We have not discussed dipole operator phenomenology
in the up quark sector.  One issue which might be of concern is the absence
of significant \dI enhancement in $D\to \pi \pi $ decays \refmark{dIDdecays}.
However, it should be noted that chromomagnetic dipole operator
coefficients in the up sector are, in general, independent of the
corresponding down sector coefficients, and could
certainly be smaller in magnitude.
A naive estimate indicates that a factor of $\sim 3$ suppression
relative to the down quark coefficients would be sufficient for transitions
between the first
two families.
This assumes that the
down quark coefficients account for $30 -50\%$ of the \dI amplitude in $K$
decays.\footnote{It is interesting to note that
attempts to solve the strong CP problem with a massless up quark
would
lead to vanishing transition dipole moments between the $u$ and $c$ quarks
because of
chiral symmetry.}

We have expressed the \dI amplitudes in terms of an
unknown ${\cal O}(p^2)$ chiral perturbation theory
suppression factor, $m_K^2 / \Lambda^2 $.
However, substantial \dI enhancement can be obtained by setting it
as low as .2, which is a reasonably conservative estimate.  Nevertheless,
theoretical progress is essential in calculating chromomagnetic
dipole matrix elements.  This is certainly also true of the
$\Delta S =2$
matrix elements, some of which play
an important role in constraining the dipole-induced \dI amplitude
in supersymmetric models.
Their `bag factors' have been crudely set to 1, according to the
vacuum insertion approximation.
We have also not taken into account leading or
next-to-leading order QCD corrections of the $\Delta S =2$
operator coefficients.

Moving to the $B$ system, in the standard model the expected inclusive
branching
ratio for
non-leptonic
charmless $b$ decays is $\approx 1-2~\%$.
On the other hand, suppression of $\B$ via chromomagnetic dipole operators
implies a
branching ratio for $b \to x g$ which is about an order of magnitude larger.
Non-leptonic charmless $b$ decays have been observed at CLEO
with a branching ratio\refmark{sharma},
\beq {\rm BR}(B^0 \to K^+ \pi^- + \pi^+ \pi^- ) =
1.8^{+0.6}_{-0.5} \pm 0.2 , \label{eq:pipi} \eeq
which is in good agreement with standard model predictions.
It is important to check that in models of $\B$ suppression the exclusive rates
associated
with
$b \to xg$ are not in conflict with this measurement.
Of course, such calculations are likely to involve considerable theoretical
uncertainty.
Ultimately, this issue should be settled by experiment.
Perhaps LEP or SLC, with their vertex detector capabilities,
could resolve the presence of charm decay vertices in non-leptonic $b$ decays
with
sufficient efficiency
to determine whether charm is not produced $15\% -30\%$ of the time.

\vglue 0.2cm

We end with a summary of our results.
In Section 3 we carried out a model-independent analysis
of dipole operator phenomenology which endeavors to study possible connections
between
\dI
enhancement, $\B$ suppression,
and the quark spectrum.
The dipole operator coefficients were therefore parametrized in terms of known
quark
masses and mixing
angles.  Our results can be classified according to the scale of flavor
physics which induces the dipole operators and quark masses.
Remarkably, there are essentially two distinct cases
in which chromomagnetic dipole operators
can lead to direct associations between \dI enhancement, or
$\B$ suppression, and observed quark masses and mixing angles.
For flavor physics in {\it{Region I}}, $M\sim 1-2~TeV$, suppression of $\B$ to
$10
-11\%$
is likely to be associated with generation of $V_{cb}$ or $m_b$.
In {\it{Region II}}, $M\sim {1\over 2 }~TeV$, the analysis suggests that
approximately
$30 -50\%$
of the observed \dI amplitude can be directly associated with
generation of $\theta_c $ or $m_s$.
$\B$ suppression can be
directly associated with generation of $V_{ub}$, or with generation of $m_b$ in
conjunction with
$V_{cb}$.
$\B$ suppression will also lead to a decrease in the
charm-multiplicity relative to the standard-model prediction, which is
consistent with recent measurements.

In Section 4 we showed that supersymmetric models
can provide explicit realizations of Region I or II
phenomenology.
In particular, for weak scale gluinos and squark masses in the $1-2~TeV$ range
it is possible to tie in $\B$
suppression with
radiative generation of $V_{cb}$.
For weak scale gluinos and $\sim {1 \over 2}~TeV$ squarks
it should be possible to obtain $\B$ suppression in
association with generation of $V_{ub}$, and $30\%
-50\%$ of the \dI amplitude
in association with generation of $m_s$ or $\theta_c$.
Some tuning between supersymmetric contributions to $\dmk$, up to 1 part in
3-4 for the larger \dI enhancements, may be required.  For larger squark masses
the
$\dmk$ constraints are weaker, but a small tuning of $\theta_c $
or $m_s$ would be required.  The most appealing scenario
arises in Region II, where all left-right down squark mass insertions
can be of same order, leading to radiative generation of $\theta_c$, $V_{ub}$,
and $m_d$,
together with \dI enhancement and $\B$ suppression.
Finally, we saw that $\B$ suppression, and even larger \dI amplitudes are
possible for
ultra-light gluinos, although a connection to the quark mass spectrum is
unlikely.
Unfortunately, in supergravity
theories with general Kahler potential, or in
string theory with moduli-driven
supersymmetry breaking, off-diagonal left-right squark
mass insertions are not large enough to obtain $\B$ suppression in Region I,
and
may
not be large enough for \dI
enhancement.

In Section 5 we discussed
an entirely different class of models in which
electroweak symmetry breaking is due to technicolor interactions, and quark
masses are
due to
techniscalar exchange \refmark{eotvos}.
We found that $\B$ suppression is possible for techniscalar masses in Regions I
and II,
and that
$30\% -50\%$ of the \dI amplitude
may be generated in Region II.
This enhancement is bounded from above by the magnitudes of
$m_s$, or $\theta_c$.  Interestingly, $\dmk$ constraints are weak
and do not play a role.
There are many possible connections between \dI enhancement, or $\B$
suppression, and
the quark
spectrum, depending on how may techniscalars
or technifermions are introduced.  This was summarized in the
previous section.
We only note that, unlike in radiative models, generation of heavy quark masses
also has rich implications for
chromomagnetic dipole operator phenomenology.
In particular, generation of $m_b$ can be associated with
$\B$ suppression at either flavor scale.
Furthermore, in Region II the top quark acquires a large chromomagnetic dipole
moment,
which would
substantially enhance the Tevatron $t \bar t$ production cross section
\refmark{rizzo}.
Finally, we investigated the effects of dimension-6 operators
on the $Z \to b \bar b$ decay width
and FCNC.
We found that $R_b$ receives substantial positive contributions,
which are correlated with contributions to $b \to s \mu^+ \mu^-$.
For example, for ${{\delta R_b} \over
R_b } \approx 3\%$, one obtains ${\rm BR} (b \to s \mu^+ \mu^- ) \sim 10^{-5}$.

Techniscalar models are appealing because
the dipole moments are
automatically tied to the quark mass spectrum.  Their magnitudes are determined
by the
techniscalar
mass(es).
An important issue for Region II phenomenology is whether techniscalar masses
as
light
as
${ 1\over 2}~TeV$ are consistent with technicolor dynamics.
We argued that this is not unreasonable, based on naive estimates of the
technifermion constituent mass.

Other models, which we did not discuss,
were also investigated.  We
found that \dI amplitudes due to dipole penguin graphs with
charged-Higgs, scalar diquarks, vectorlike quarks, or leptoquarks in the
loop tend to be smaller, because of more restrictive $\dmk$ constraints.
In fact, it is difficult to find models which can match the
dipole-induced \dI
enhancement possible in supersymmetric and
techniscalar models.
Models of quark substructure are potential candidates \refmark{peskin},
since they are likely to produce transition dipole moments in association
with quark mass
generation, but a fairly
light compositeness scale would be required.

We end with implications of $\B$ suppression for radiative $B$ decays.
A general model-independent criterion, applied at the scale of new flavor
physics,
distinguishes those
models of $\B$ suppression which do not conflict with the inclusive measurement
of ${\rm
BR}(b \to x
\gamma )$. We have seen that it can be applied to a rather general
class of models
with new scalar bosons at the TeV scale.
The analysis also suggests that those models which survive in Region II
may produce {\it{large rates for}} $b \to d \gamma$.
The corresponding branching ratio would lie in the range $10^{-5} - 10^{-
4}$, which is substantially
larger than the standard model prediction.   This is, in fact, the case in both
the
supersymmetric and
techniscalar models we have studied.
Implications for $B \to \rho \gamma$,
or $B \to \omega \gamma$
offer another example of the richness of Region II phenomenology.
Finally, our main result can
be summarized by
comparing the model-independent, supersymmetric, and techniscalar
plots of Figs. 4, 8, and 12, respectively.
Their obvious similarity strongly suggests that substantial \dI enhancement,
$\B$ suppression, and the quark mass
spectrum are tied together by chromomagnetic dipole operators which are
induced by new flavor physics at the TeV scale.

\vskip.25in
\centerline{ACKNOWLEDGEMENTS}

It is a pleasure to thank Michael Peskin for useful and
crucial discussions, and for helpful comments regarding the manuscript.
I would also like to thank John Donoghue for a discussion
of \dI dipole operator matrix elements, and
Michael Dine, Adam Falk, Matthias
Neubert, Yossi Nir and Michal Porat for useful conversations. Thanks are also
due to
Joanne Hewett, Tom Rizzo, Sheldon Stone and Steve Wagoner for help with
experimental issues.

\newpage

{\elevenbf\noindent Appendix A.   More on Connections to the Quark Mass
Spectrum
}
\vglue 0.2cm

Throughout this paper we have expressed the physical transition dipole
operators in terms of partial contributions to the down quark mass
matrix, in the mass eigenstate basis.
In order to study the connection between these contributions and
known features of the quark spectrum it is necessary to reexpress them in terms
of mass
contributions in
the quark interaction basis.
This task is simplified if we make some reasonably general assumptions about
the
form of
the full down
quark mass matrix, $M^d$, after all individual contributions have been taken
into account.
In particular, we always assume that an interaction basis exists in which
the entries $M^d_{i j} \bar{d}^i_L d^j_R $
satisfy the hierarchy
\begin{eqnarray*}
M^d=\left(\matrix{\sim m_d & \sim m_d & \sim m_d \cr
\sim m_d & \sim m_s & \sim m_s \cr \sim m_d & \sim m_s & \sim m_b
}\right),~~~~~~~~~~~~~~~~~~~~~~~~~~~~~~~~~~~~~~~~~~~~~~~~~~~~~(A1)
\end{eqnarray*}
with similar assumptions for the up quark matrix.
Eq. (A1) is intended to be schematic.  For example, the $(12)$ entry will
actually be $\approx \theta_c m_s$ which is several times larger than $m_d$.
Given eq. (A1) and it's analogue for the up sector,
the KM angles will essentially be generated in the down sector.\footnote{The
$(32)$ and
$(31)$ entries in eq. (A1) are
unrelated to the KM angles and, in general, can be as large as a few GeV.
However, in this case the connection between quark masses, or mixing angles,
and $\B$ suppression is lost.}

Diagonalization of eq. (A1) is straightforward.
The down quark masses are given by
\begin{eqnarray*}
 m_d \approx m_{11}-{m_{12} m_{21} \over m_s},~~m_s \approx m_{22}
-{m_{23} m_{32} \over m_b},~~m_b \approx m_{33}.~~~~~~~~~~~~~~~~~~~~~~~~(A2)
\end{eqnarray*}
The down quark mass eigenstates are given in terms of the interaction
basis by
\begin{eqnarray*}
\vert \psi^i_L \rangle = x_{ij}^L \vert d_L^j \rangle,
{}~~~~\vert \psi^i_R \rangle=  x_{ij}^R
\vert d_R^j \rangle,~~~~~~~~~~~~~~~~~~~~~~~~~~~~~~~~~~~~~~~~~~~~~~~~~~~~~(A3)
\end{eqnarray*}
where $  \psi^1_L$, ${\psi}^2_L$, ${\psi}^3_L$ are the left-handed quark
mass eigensates $d_L$, $s_L$, $b_L$, respectively, and $R$ subscripts label the
corresponding right-handed quarks.
Taking all parameters to be real, the $x^{ij}_L$ are given by
\begin{eqnarray*}
x_{ii}^L \approx 1,~~~x_{21}^L \approx {m_{12} \over m_{22}} \approx V_{us} ,
{}~~~x_{32}^L \approx
{m_{23} \over m_{33}} \approx V_{cb} ,
{}~~~x_{31}^L \approx
{m_{13}\over m_{33}} \approx V_{ub},
\end{eqnarray*}
\begin{eqnarray*}
 x_{12}^L \approx -x_{21}^L , ~~~x_{23}^L \approx -x_{32}^L ,
{}~~~x_{13}^L \approx -{m_{13} \over m_b} +{m_{23} \over m_{33}}
{m_{12} \over m_{22}} \approx V_{td}.~~~~~~~~~~~~~~~~~~~~(A4)
\end{eqnarray*}
Expressions for the $x_{ij}^R $
are obtained from the above by interchanging indices on the $m_{ij}$.
The up quark masses and
eigenstates are completely analogous.
Note that
\begin{eqnarray*}
 V_{ts} \approx - V_{cb}, ~~~  V_{td} \approx -V_{ub} + V_{cb} V_{us}
{}~~~~~~~~~~~~~~~~~~~~~~~~~~~~~~~~~~~~~~~~~~~~~~~~~~~(A5)
\end{eqnarray*}
in the limit of a diagonal up matrix.

We can now investigate claims made in Section 3 about the correspondence
between
certain
ranges for the off-diagonal down quark mass matrix contributions, in the mass
eigenstate basis,
and generation of
observed features of the quark spectrum.
Recall that the parametrization of Section 3, eq. (24),
corresponds to generation of rank-1 dipole operator coefficient matrices,
$C_G^{ij}$, and proportional
rank-1 mass matrices
${\scriptstyle \Delta} m_{ij} $.
In general, there may be several such contributions,
each one generated by a different exchange of particles.
We assume that
these do not upset the hierarchy in (A1), so that large cancelations
among different contributions to the
mass matrix are not required.

\vglue 0.2cm
{\noindent \bf{(a)} }  $\abs{\dmdsp} \sim \abs{\theta_c m_s}$ can
be associated with generation of $\theta_c$, or $m_s$, but not both.
This can be seen by expressing $\dmdsp$ in the interaction basis,
\begin{eqnarray*}
\dmdsp = {\scriptstyle \Delta} m_{12} -  {m_{12} \over m_{22}}
{\scriptstyle \Delta} m_{22} + {\cal O}
(\theta_c m_d).~~~~~~~~~~~~~~~~~~~~~~~~~~~~~~~~~~~~~~~~~~~~~~~~~(A6)
\end{eqnarray*}
If ${\scriptstyle \Delta} m_{12}$ accounts for the bulk of $m_{12}$
then the induced mass matrix generates $\theta_c$.  Alternatively, if
${\scriptstyle \Delta} m_{22}$ accounts for the bulk of $m_{22}$,
then it generates $m_s$.  In either case one
obtains $\abs{\dmdsp} \sim \abs{\theta_c m_s}$.  However, the
limit in which
both $\theta_c$
and $m_s $ are generated by ${\scriptstyle \Delta} m_{ij}$ leads to
suppression of
$\abs{\dmdsp}$.

\vglue 0.2cm
{\noindent \bf{(b)} }  $\abs{\dmsbp} \sim \abs{V_{cb} m_b }$
can be associated with generation of $V_{cb}$, or $m_b$, but
not both.
In the interaction basis
\begin{eqnarray*}
\dmsbp = {\scriptstyle \Delta} m_{23} -
{m_{23} \over m_{33}} {\scriptstyle \Delta} m_{33} + {\cal O}
(V_{cb}  m_s).~~~~~~~~~~~~~~~~~~~~~~~~~~~~~~~~~~~~~~~~~~~~~~~(A7)
\end{eqnarray*}
If ${\scriptstyle \Delta} m_{23}$ accounts for the bulk of $m_{23}$
then the induced mass matrix generates $V_{cb}$.  Alternatively, if
${\scriptstyle \Delta} m_{33}$ accounts for the bulk of $m_{33}$,
then the induced mass matrix generates $m_b$.  In either case one obtains
$\abs{\dmsbp}
\sim \abs{V_{cb} m_b}$.  However,
if both $V_{cb}$
and $m_b $ are generated by ${\scriptstyle \Delta} m_{ij}$, then $\abs{\dmsbp}$
is suppressed.
For example, if
\begin{eqnarray*}
m_{33} -{\scriptstyle \Delta} m_{33} \sim V_{cb} m_b,~~~
m_{23}-{\scriptstyle \Delta} m_{23} \sim  V_{ub}
m_b,~~~~~~~~~~~~~~~~~~~~~~~~~~~~~~~~~~~(A8)
\end{eqnarray*}
then $\abs{\dmsbp} \sim \abs{V_{ub} m_b}$.

\vglue 0.2cm
{\noindent \bf{(c)} }  $\abs{\dmdbp} \sim \abs{V_{ub} m_b }$
can be associated with generation of $V_{ub}$, $V_{cb}$, or
$m_b$, but not all three.
In the interaction basis
\begin{eqnarray*}
\dmdbp = {\scriptstyle \Delta} m_{13}
-{m_{12} \over m_{22}} {\scriptstyle \Delta} m_{23}
-{m_{13} \over m_{33}} {\scriptstyle \Delta} m_{33} +{\cal O}
(V_{cb} V_{ub} m_b)~~~~~~~~~~~~~~~~~~~~~~~~~~~~~(A9)
 \end{eqnarray*}
If ${\scriptstyle \Delta} m_{ij}$ generates $V_{ub}$, $V_{cb}$,
or $m_b$ then it must account for the bulk of $m_{13}$, $m_{23}$,
or $m_{33}$, respectively.
Clearly, if any one, or any two of these possibilities is true, one obtains
$\abs{\dmdbp} \sim \abs{V_{ub} m_b }$.  However, in the limit that all three
are
true,
$\abs{\dmdbp}$ is much smaller.

It is clear from (b) and (c) that $\xi' \sim .1$ (see eq.
(\ref{eq:xitenth}))
can correspond to
generation of $V_{cb}$ in conjunction with $m_b$, or to generation of
$V_{ub}$, whereas $\xi' \sim 1$ (see eq. (\ref{eq:xione})) can correspond to
generation of
$V_{cb}$ or $m_b $, but not both.

\vglue 0.4cm
{\noindent \it Supersymmetry}
\vglue 0.2cm

We saw that in supersymmetric models ${\scriptstyle \Delta} m_{ij}$ is
generally rank-3, but still proportional to the dipole moment matrices,
up to very small corrections.
The only change to the above analysis is that radiative
generation of $m_b$ is not an option.  This means, in particular, that
Region II scenarios of $\B$ suppression can be associated with
generation of $V_{ub}$, but not $V_{cb}$.

An issue of relevance for \dI enhancement with {\it{ultra-light
gluinos}} is
how large a hierarchy between $\delta \tm^2_{d_L s_R} $ and
$\delta \tm^2_{d_R s_L} $ is possible, for the purposes of
evading $\dmk$
constraints.
According to Figs. 6 and 9,
large enhancement requires a hierarchy of ${\cal O}(30)$
for weak scale squarks, and ${\cal O}(300)$ for squarks near a TeV.
To settle this issue it is useful to express the squark mass insertions in the
quark
interaction basis,
\begin{eqnarray*}
  \delta \tm^2_{d_L s_R} \approx  \delta \tm^2_{1_L 2_R}
+ x_{12}^L \delta \tm^2_{2_L 2_R}+ x_{21}^R  \delta \tm^2_{1_L 1_R}+
 x_{12}^L x_{21}^R \delta \tm^2_{2_L 1_R}+...
 \end{eqnarray*}
 \begin{eqnarray*}
 \delta \tm^2_{d_R s_L} \approx  \delta \tm^2_{2_L 1_R} +
+ x_{12}^R \delta \tm^2_{2_L 2_R}+ x_{21}^L \delta \tm^2_{1_L 1_R}+
 x_{12}^R x_{21}^L \delta \tm^2_{1_L 2_R}+... ~~~~~~~~~~(A10)
 \end{eqnarray*}
Terms involving 3rd generation squark mass insertions have not been shown
explicitly.
Given eq. (A1), a hierarchy of ${\cal O}(30)$ requires
a similar hierarchy between $\delta \tm^2_{1_L 2_R}$ and $\delta
\tm^2_{2_L 1_R}$, and an order of magnitude hierarchy
between $\delta \tm^2_{1_L 2_R}$, and both $\delta
\tm^2_{2_L 2_R}$ and $\delta \tm^2_{1_L 1_R}$.
So in scenarios with ultra-light gluinos and weak scale squarks
several non-trivial conditions must be satisfied.
Assuming that the bulk of $\theta_c$ is generated in the down sector,
an upper bound on
the ratio of $\delta \tm^2_{d_L s_R} $ to
$\delta \tm^2_{d_R s_L} $ of ${\cal O}(400)$, corresponding to
$(x_{12}^R x_{21}^L)^{-1}$,
is obtained by setting all squark mass
insertions to zero except $\delta \tm^2_{1_L 2_R}$.
We have used a lower bound for
$x_{12}^R $ of ${\cal O} ( {m_d \over m_s}\theta_c)$, obtained in the
limit that the
$m_{21}$ entry of the down quark mass matrix vanishes.
Strictly speaking, scenarios with ultra-light gluinos
and squarks near a TeV are possible, but they are
highly constrained and clearly
disfavored, even though they may lead to
generation of $\theta_c$, as
noted in Section 4.

\vglue 0.4cm
{\noindent \it Techniscalar models}
\vglue 0.2cm

In techniscalar models in which both third and second generation masses
are due to techniscalar exchange the down quark mass matrix is generally
of the form
\begin{eqnarray*}
 M^d = m_3 \vert h^3_L \rangle \langle h^3_R \vert + m_2 \vert
h^2_L \rangle \langle h^2_R \vert + \delta M.
{}~~~~~~~~~~~~~~~~~~~~~~~~~~~~~~~~~~~~~~~~~~~~~(A11)
\end{eqnarray*}
The bras and kets are dimensionless 3-component vectors, normalized to
unity, obtained from Yukawa coupling vectors
like ${\bf h}$, ${\bf h}^d$ in (\ref{eq:tcmasses}).
The massive coefficients have magnitudes $m_3 \sim m_b $, $m_2 \sim
m_s$, and the matrix $\delta M$ is generally rank 3 or
less,
with entries which are typically ${\cal O}(m_d)$, or ${\cal O}(\theta_c
m_s)$. For example, if 1st generation masses and mixing angles are also
due to techniscalar exchange then $\delta M$ is rank 1, however, if
they are due to some radiative mechanism then $\delta M$ might be rank 3.
The up quark matrix is of the same form. $SU(2)_L$
implies that $\vert h^3_L \rangle $ and $\vert h^2_L \rangle$
are equal for the up and down matrices,\footnote{For
two copies of the technifermions $T_R$, $U_L$, $D_L$ and a single
techniscalar, this also assumes that the up and down
condensates
respect custodial isospin symmetry which, of course, is also required
by the $\rho$ parameter constraint.}
which insures the near alignment of up and down mass eigenstates
required by the KM mixing hierarchy.

It is easy to show that the following interaction basis
reproduces the hierarchy in (A1) for $M^d$:
\begin{eqnarray*}
  \vert 3_L \rangle = \vert h^3_L \rangle,~~~
\vert 2_L \rangle  \propto \vert h^2_L \rangle - \vert h^3_L \rangle
\langle h^3_L \vert h^2_L \rangle,~~~\vert 1_L \rangle \perp \vert 2_L
\rangle,~\vert 3_L \rangle.~~~~~~~~~~~~~~~~(A12)
\end{eqnarray*}
The right-handed basis elements are completely analogous.
Note, in particular, that the first two terms in (A11)
correspond to the lower right $2\times 2$ submatrix of eq. (A1)
up to corrections due to $\delta M$. This explains item (b)
at the end of Section 5, since with two sets of technifermions ($N_D
=2$)
and a single techniscalar
the dipole operator coefficient matrices are proportional to the sum of the
first two terms. This means that transition dipole moments must be
proportional to matrix elements of $\delta M$, leading to Region II
phenomenology.
If only 3rd generation masses are due to techniscalar exchange
then $M^d$ can still be written in the form of
(A10), but only the
first term would be due to techniscalar exchange,
accounting for the bulk of the (33) entry in eq.(A1).
This explains item (a), since the dipole
coefficients would be proporional to the first term in eq.
(A10), leading to Region I phenomenology.
Similarly, in the case of item (c) the
dipole coefficient matrices correspond to a sum of two distinct contributions,
proportional to the first and second terms in (A10), respectively.
In the absence of substantial accidental cancelations,
each of these contributions leads to Region I phenomenology.
Finally, in (d) the dipole coefficients are proportional to $\delta M$,
again leading to Region II phenomenology.

\vglue 0.6cm
{\elevenbf\noindent Appendix B.  New contributions to $\dmk$ and
$\dmB$    }
\vglue 0.2cm

{\noindent\it Supersymmetry}
\vglue 0.2cm
The supersymmetric contributions to $\dmk$ are given by\footnote{
The signs of all terms which include the enhancement factor $R_K$ are opposite
to those in \refmark{kelley}. The source of the
discrepancy is in the vacuum insertion
matrix elements which have been used.
Our matrix elements are consistent with ref. \refmark{trampetic}.}
\refmark{kelley}
\begin{eqnarray*}
 \dmk = { \alpha_s^2 \over {216 m_{\tq}^2 }}\left({2 \over 3}f_K^2
 m_K \right)
\Bigg[\left({{\delta \tm^4_{d_L s_R}} \over m_{\tq}^4 }\right)
216 R_K x f_6 (x)  \end{eqnarray*}
\begin{eqnarray*}~~~~~~~~+\left({{\delta \tm^4_{d_R s_L}} \over m_{\tq}^4
}\right)
216 R_K x f_6 (x)
+\left( {{\delta \tm^2_{d_L s_R}} \over m_{\tq}^2 }
{{\delta \tm^2_{d_R s_L}} \over m_{\tq}^2 }\right) 108 \tilde{f}_6 (x)
\end{eqnarray*}
\begin{eqnarray*}~~~~~~~~-\left( {{\delta \tm^4_{d_L s_L}} \over m_{\tq}^4 }+
{{\delta \tm^4_{d_R s_R}} \over
m_{\tq}^4 }\right)
(66 \tilde{f}_6 (x) +24 x f_6 (x) ) \end{eqnarray*}
\begin{eqnarray*}~~~~~~~~ +\left({{\delta \tm^2_{d_L s_L}} \over m_{\tq}^2 }
{{\delta \tm^2_{d_R s_R}} \over m_{\tq}^2}\right)
\left([-36 +24 R_K] \tilde{f}_6 (x) -[72 +384 R_K ]x f_6 (x) \right)
\Bigg], ~~~~~~(B1)
\end{eqnarray*}
where $f_K = 161~MeV$, $x = {m_{\tg}^2 /  m_{\tq}^2}$,
\begin{eqnarray*}
 R_K \equiv  \left({m_K \over {m_s +m_d}}\right)^2 ,
\end{eqnarray*}
and
\begin{eqnarray*}
 f_6 (x) = {1 \over {6(1-x)^5}} (-6 {\rm ln} x -18x {\rm ln} x -x^3
+9x^2+9x-17)
\\
 \tilde{f}_6 (x) ={ 1 \over {3(1-x)^5}}(-6 x^2 {\rm ln} x -6 x {\rm ln} x +x^3
+9x^2 - 9 x -1).
\end{eqnarray*}
We have used the vacuum
insertion approximation for all matrix elements, with
$m_s =150~MeV$ and $m_d = 8~MeV$.
$\alpha_s $, $m_{\tg}$ and $m_{\tq}$ are taken at the squark mass scale, and
QCD
corrections are not included.
Supersymmetric contributions to $\dmB$ in the vacuum insertion
approximation are obtained from
(B1) via the appropriate flavor substitutions.
We have taken $f_B \approx 230~MeV$,
and $m_b  = 4.25~GeV$ in $R_B$, corresponding to the running mass $m_b
(m_b)$.

\vglue 0.4cm
{\noindent\it Techniscalar models}
\vglue 0.2cm

We give a crude estimate for the contributions to $\dmk$
of box graphs with techniscalars and technifermions in the loop in
the vacuum insertion approximation, without QCD corrections.
The technifermions are assigned a mass $m_{TC}$, equal to
the `constituent' mass in
eq. (\ref{eq:mTC}). The simplest case is considered, corresponding to
$N_D=1$, and exchange of a single
techniscalar $\om$.
The $N_D =2$ case is slightly more restrictive, but
our conclusions do not change significantly.
We obtain
\begin{eqnarray*}
 \dmk = {{N_{TC} m_K f_K^2} \over 12} \Bigg[
 \left({{\lambda^2_d \overline{\lambda}^2_s }\over 2}
 +{{\overline{\lambda}^2_d {\lambda}^2_s }\over 2}\right) {z^2 \over m_{TC}^2}
I(z) R_K +
(\lambda_d^2 \lambda_s^2 + \overline{\lambda}_d^2 \overline{\lambda}_s^2 )
{\tilde{I}(z)\over m_{\om}^2}  \end{eqnarray*}
\begin{eqnarray*}
{}~~~~~~~~~~+\lambda_d \lambda_s \overline{\lambda}_d \overline{\lambda}_s
\left({z \over
m_{\om}^2} I(z) \left[2 R_K +3 \right] - {\tilde{I}(z) \over m_{\om}^2}
\left[{3 \over 2} R_K +{1 \over 4} \right] \right)\Bigg] ,
{}~~~~~~~~~~~~~~~~~~~~(B2)
\end{eqnarray*}
where $z = {m_{TC}^2 /  m_{\om}^2}$, and
\begin{eqnarray*}
 I(z)= {1 \over {16 \pi^2}} {{-2 +2z-(1+z) {\rm ln} z} \over (1- z)^3 }
\end{eqnarray*}
\begin{eqnarray*}
\tilde{I}(z) =  {1 \over {16 \pi^2}}{{z^2 -1-2z {\rm ln} z}\over (1-z)^3}.
\end{eqnarray*}

\bigskip

\newpage
\renewcommand{\baselinestretch}{1}

\newpage
{\bf Figure Captions}
\begin{itemize}
\item[Figure 1.] {Parton-model predictions for $\B$ versus
$\alpha_s (M_z)$, evaluated at $\mu =m_b$.
(a) $m_b =4.9~GeV$, $m_c =1.4~GeV$, (b) $m_b =4.8~GeV$,
$m_c =1.4~GeV$, (c) $m_b =4.7~GeV$, $m_c =1.35~GeV$,
(d) $m_b =4.6~GeV$, $m_c = 1.2~GeV$.}
\item[Figure 2.] {Contours of $\B=.11$ (solid), $.10$ (dashed) in the
plane of $\abs{\zeta_G \xi'}$ vs. $M$ for $m_t =170~GeV$, $m_b
=4.8~GeV$,
$m_c=1.4~GeV$,
and $\Lambda^{(4)} =300~MeV$. }
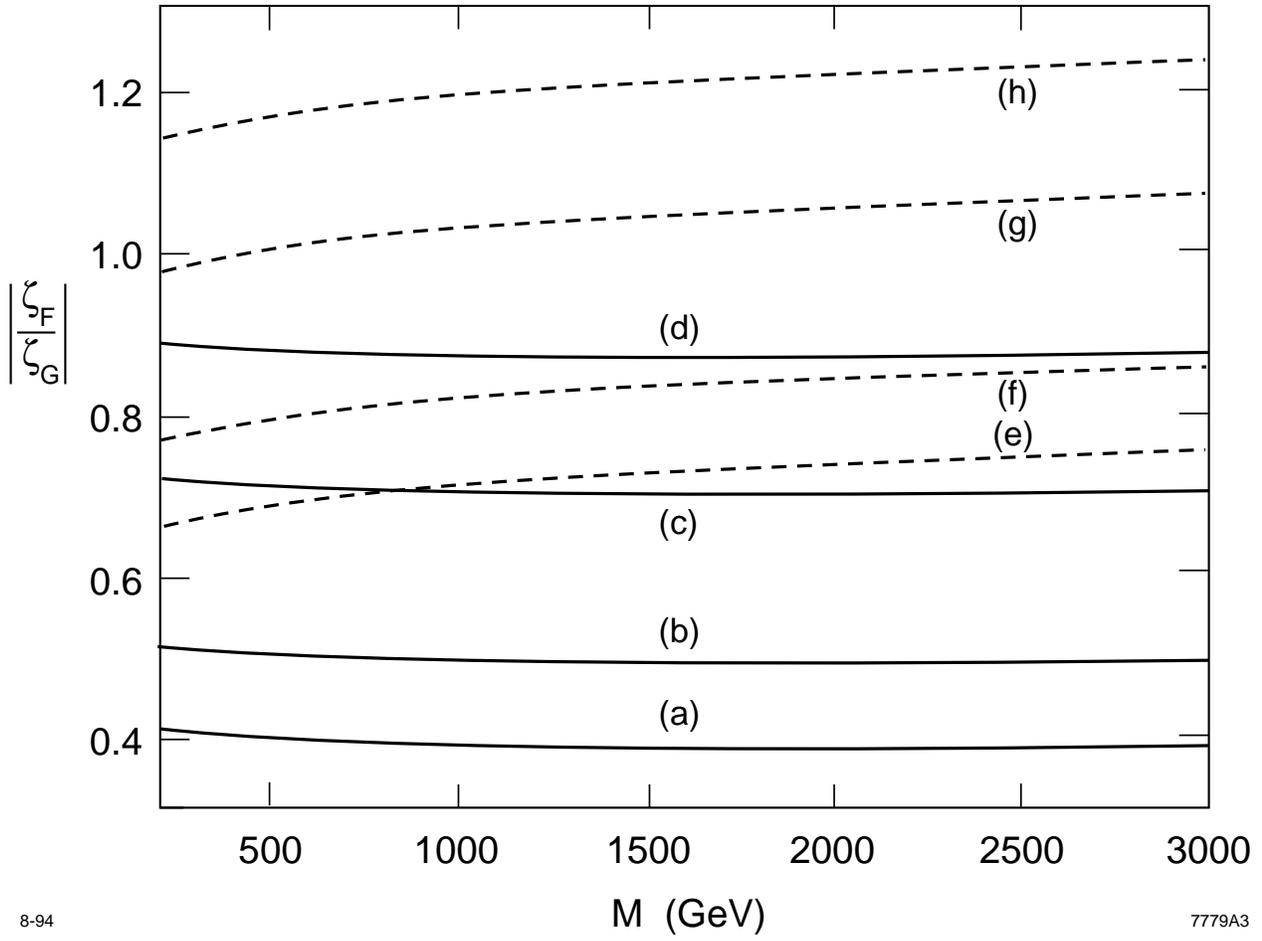
\item[Figure 3.] {The ratio $\zeta_F \over \zeta_G$ vs. $M$,
for different values of $\B$,
${\rm BR}(b\to x \gamma)$, and ${\rm sign}({\zeta_F}) = -{\rm
sign}({\zeta_G})$
(solid lines), ${\rm sign}( {\zeta_F}) = {\rm sign}( {\zeta_G})$ (dashed
lines).
Curves (a),(b),(e),(f) and (c),(d),(g),(h) correspond to ${\rm BR}(b\to x
\gamma)= 2
\times 10^{-4}$ and
$4\times 10^{-4}$, respectively.
$\B=.10$ in (a),(c),(e),(g) and .11 in (b),(d),(f),(h).  $m_b =4.8~GeV$, $m_t
=170~GeV$, and
$\Lambda^{(4)} =300~MeV$.}
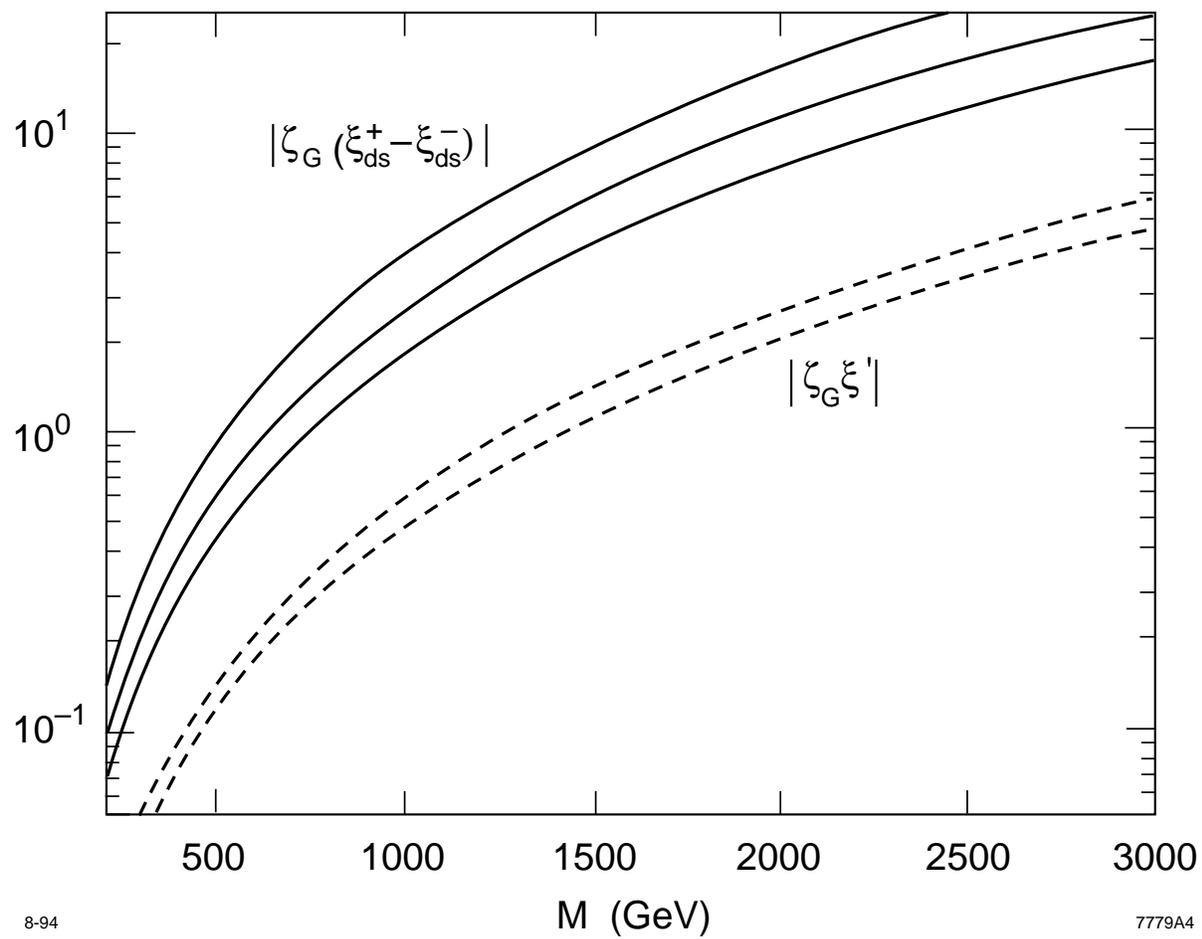
\item[Figure 4.] {Contours, from top to bottom, of $R_0 =(1.5,1.0,.7){m_K^2
\over
\Lambda^2}$ (solid
curves),
in the plane of $\abs{\zeta_G (\xi_{ds}^+ - \xi_{ds}^-)}$ vs. $M$,
and contours of $\B =.10,.11$ (dashed) in the plane of $\abs{\zeta_G
\xi'}$ vs. $M$ plane. We've taken
$m_s (m_c)
=150~MeV$, $m_c =1.4~GeV$,
$m_b = 4.8~GeV$, $m_t =170~GeV$ and $\Lambda^{(4)} =300~MeV$.}
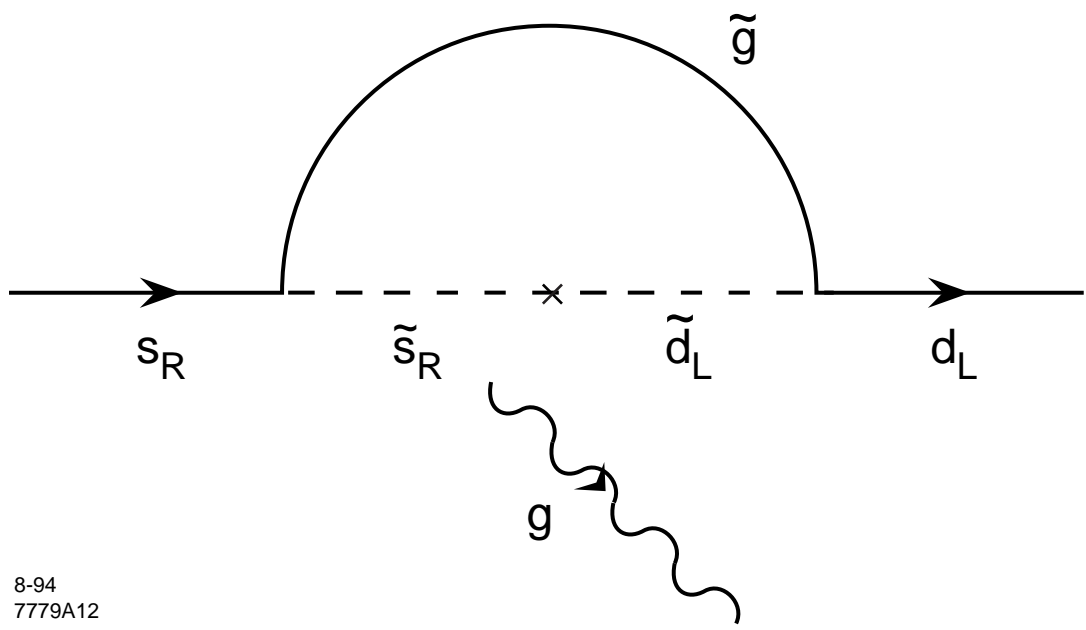
\item[Figure 5.] {Gluino penguin graphs giving rise to
chromomagnetic transition dipole moments.
The gluon is attached in all possible ways.}
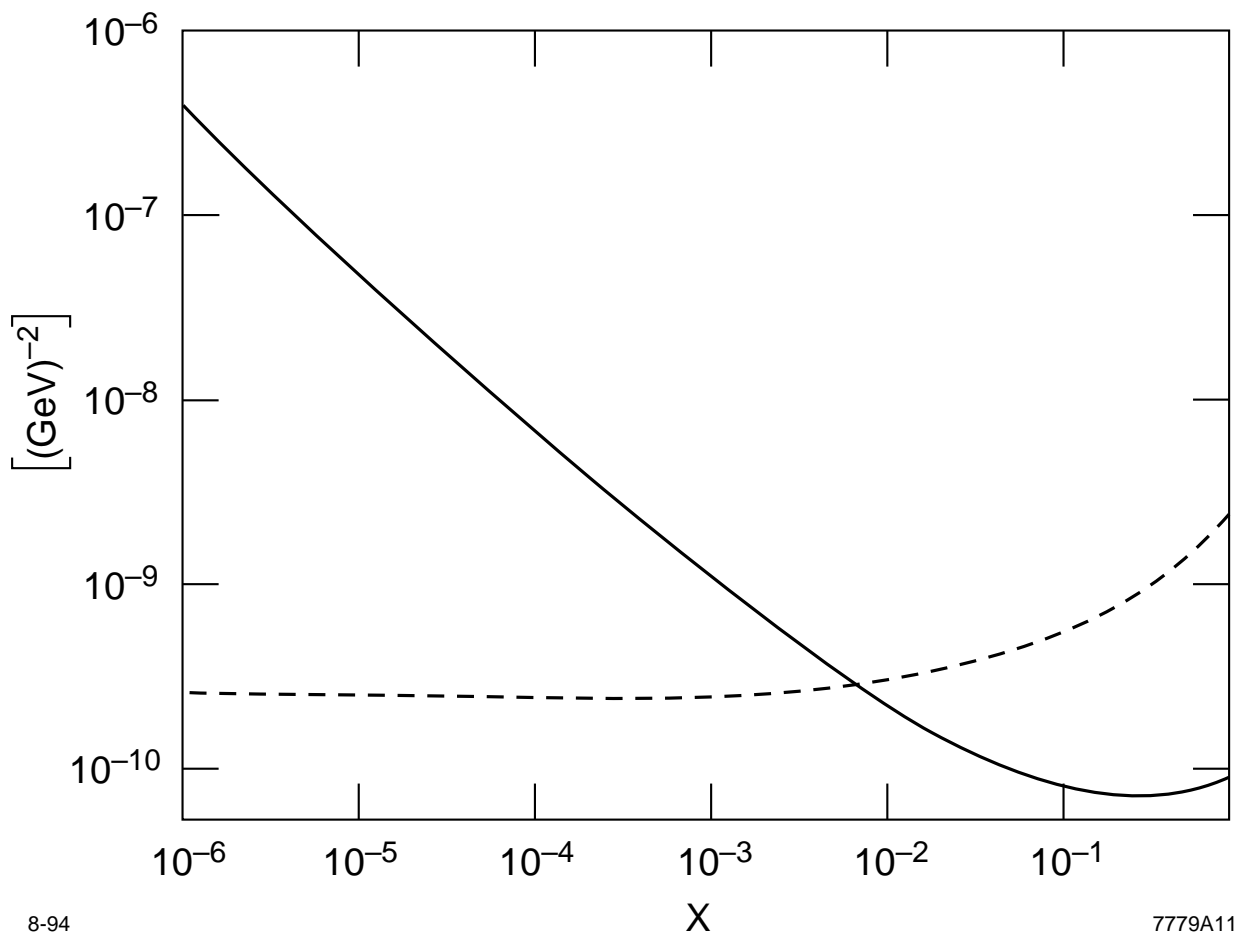
\item[Figure 6.] {Upper bounds on ${\d
\tm^4_{d_L s_R} \over
m_{\tq}^6}$ (solid) and ${{\d \tm^2_{d_L s_R} \d \tm^2_{d_R s_L}} \over
m_{\tq}^6}$ (dashed ) from gluino box-graph
contributions to $\dmk$.}
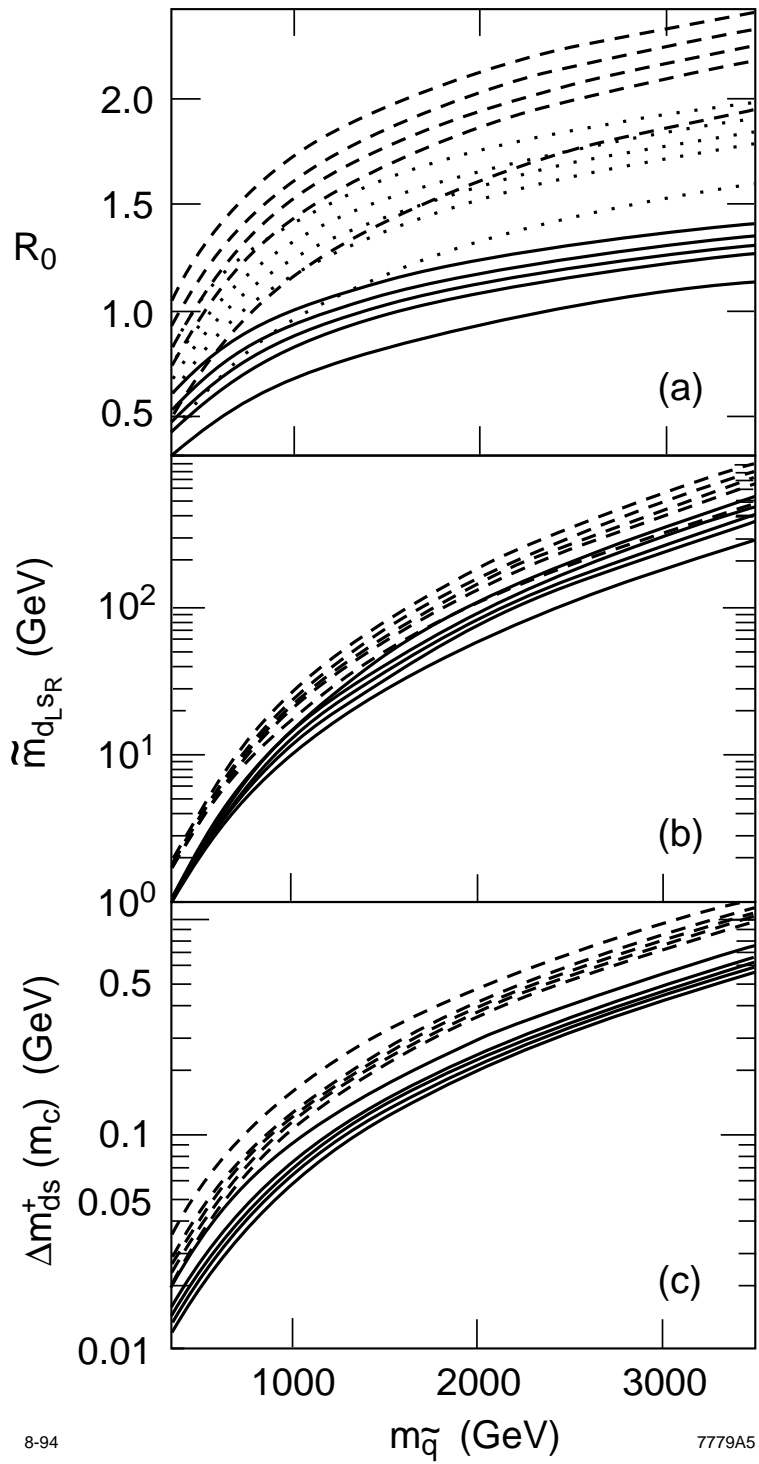
\item[Figure 7.] {(a) Contours of
$\dmk=\dmk^{exp}$ (solid ),
$2 \dmk^{exp}$ (dotted), $3 \dmk^{exp}$ (dashed) in the
plane of $R_0$ vs. $m_\tq$.  In each case,
$m_{\tg} (m_{\tg}) =125,150,175,200$,$300~GeV$,
from top to bottom. $\Lambda^{(4)} =300~MeV$, $m_t =170~GeV$,
$m_b =4.8~GeV$, $m_c = 1.4~GeV$ and $m_s (m_c) =150~MeV$.
(b) Contours of $\dmk=\dmk^{exp} $ (solid),
$3 \dmk^{exp}$ (dashed) in the plane of
$\tilde{m}_{d_L s_R}(m_{\tq})$ vs.$m_{\tq}$. The gluino masses increase
from top to
bottom. (c) Contours of $\dmk= \dmk^{exp}$ (solid),
$3 \dmk^{exp}$ (dashed) in the plane of $\dmdsp$ vs. $m_{\tq}$.  Gluino masses
decrease from
top to bottom.}
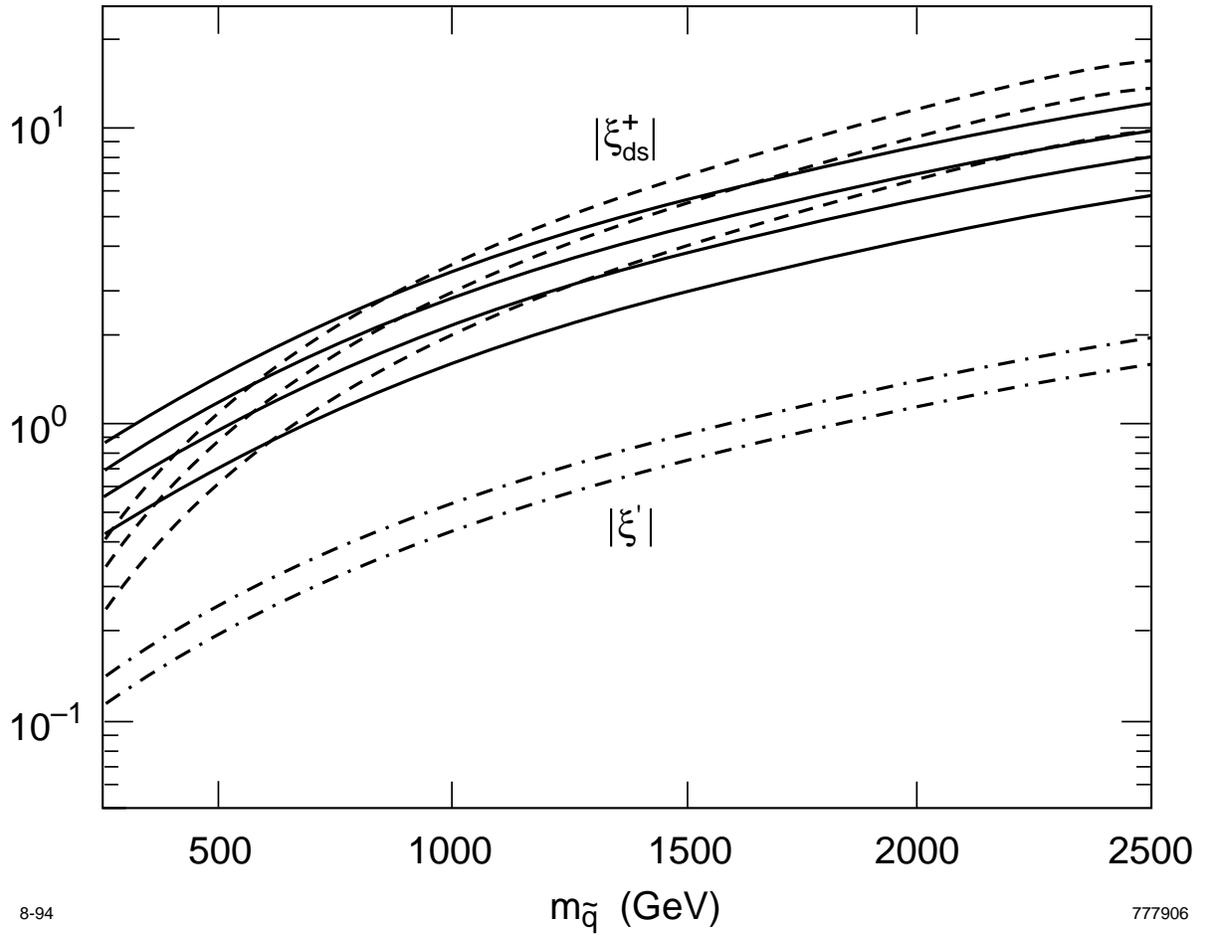
\item[Figure 8.] {Contours of $R_0 =(1.5,1.25,1.0,.75){m_K^2 \over \Lambda^2}$
(solid)
and $\dmk = 3\dmk^{exp}$, $2 \dmk^{exp}$, $\dmk^{exp}$ (dashed) in the
plane of $\xi_{ds}^+$ vs. $m_{\tq} $.
$R_0$ and $\dmk$ decrease from top to bottom.
Also included are contours of $\B=.10,.11$ in the
plane of $\xi'$ vs. $m_\tq $ (dot-dashed).  The quark mass thresholds and
$\Lambda^{(4)}$ are
as in Fig. 7.}
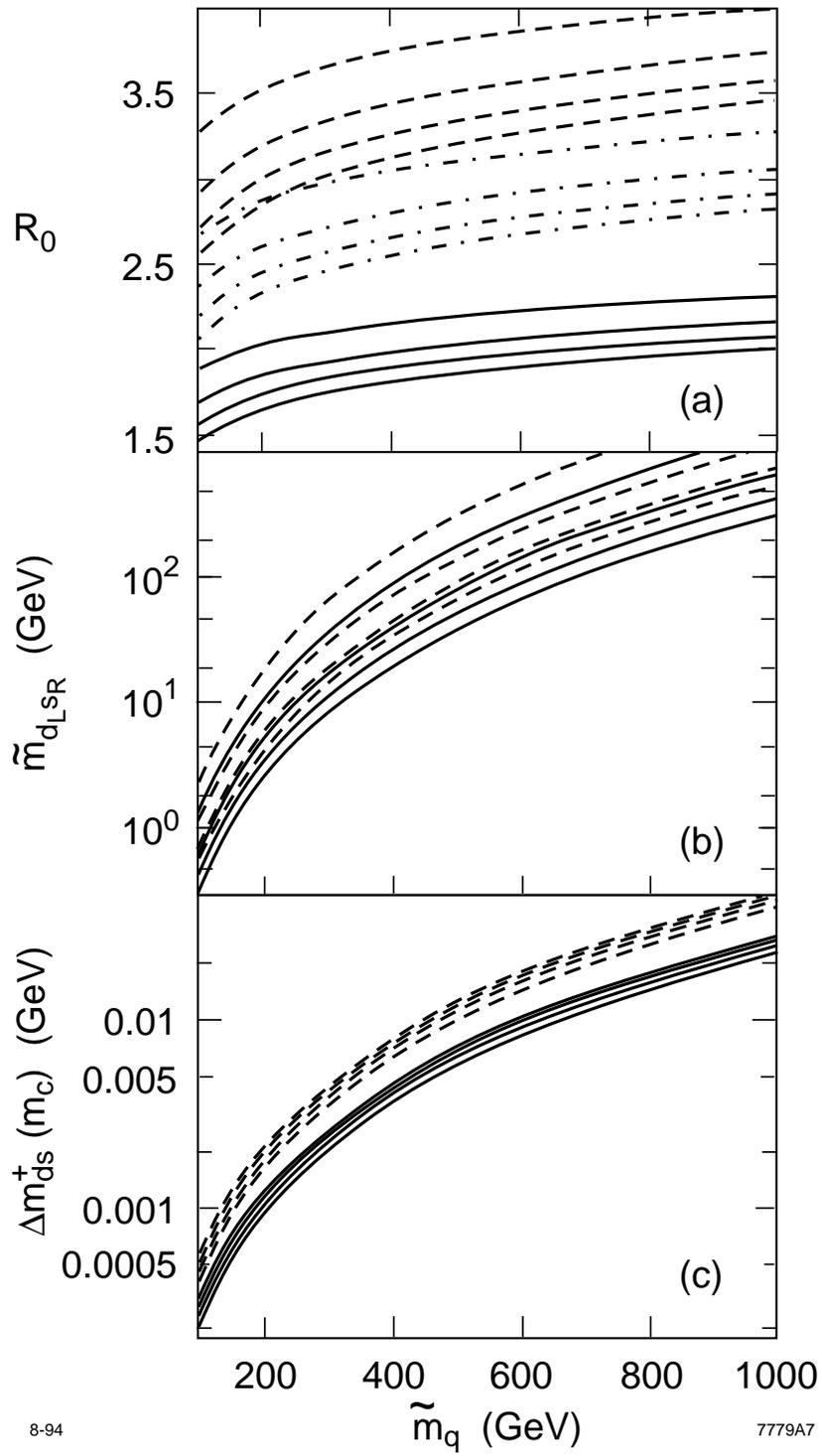
\item[Figure 9.] {Figs. 9a,b,c are the same as Figs. 7a,b,c, but for $m_{\tg}
=1,2,3,4~GeV$.
Gluino masses
increase from top to bottom
for each value of $\dmk$ in 9a, 9b, but decrease in 9c.
Evolution from $m_{\tq}$ to $m_c$ is for
$\alpha_s (M_z) = .124$, and the usual quark mass thresholds.}
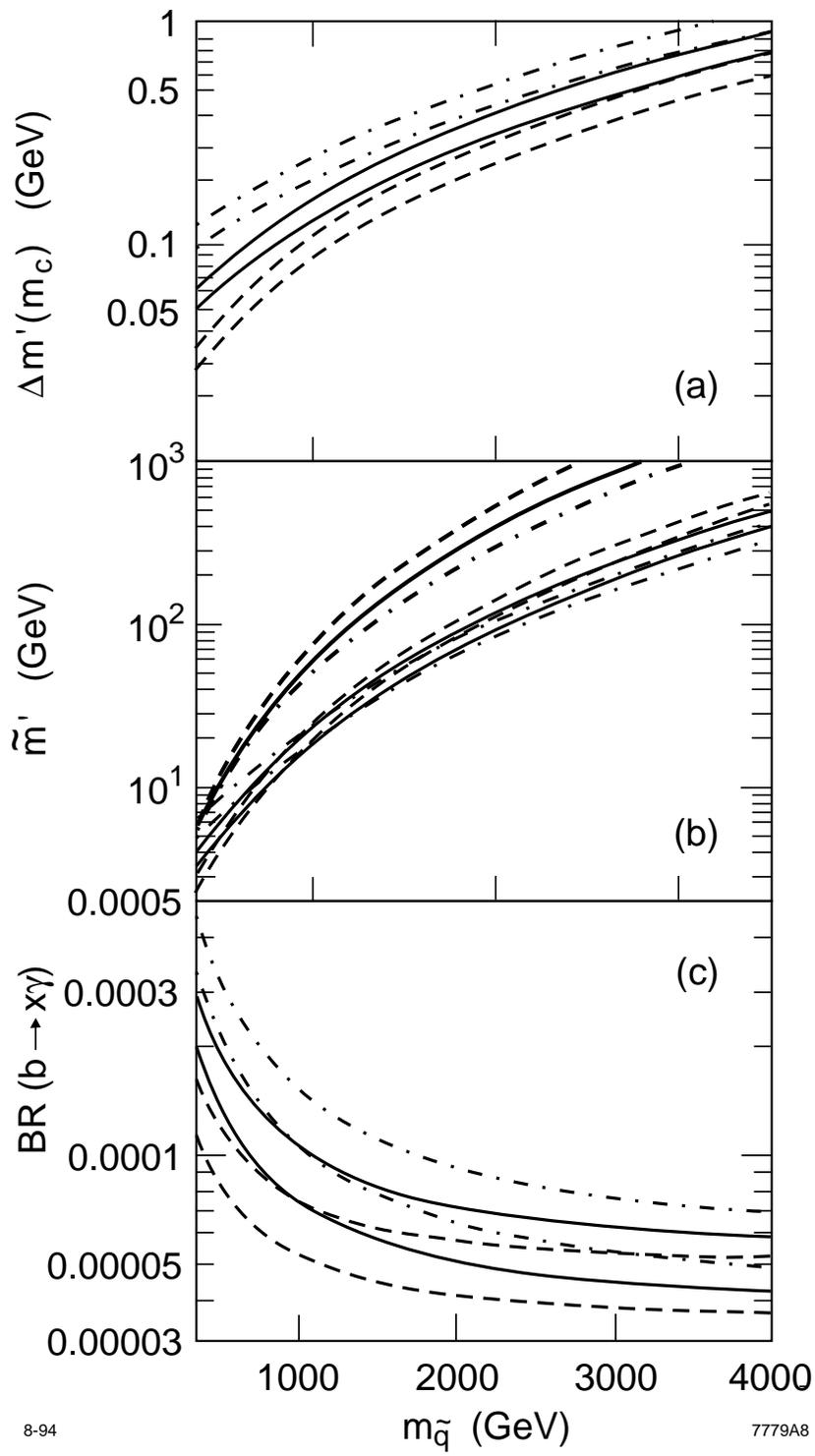
\item[Figure 10.] {(a) Contours of $\B = .10,.11$
in the plane of ${\scriptstyle \Delta}  m' (m_c)$ vs. $m_{\tq }$.
(b)  Contours of $\B = .10,.11$
in the plane of $\tilde {m}' (m_{\tq})$ vs. $m_{\tq}$,
together with upper bounds (thick curves) on
$\tilde{m}_{d_L b_R} (m_{\tq} )$ from $\dmB$. (c) Contours
of
$\B = .10, .11$
in the plane of ${\rm BR}(b \to x \gamma)$ vs. $m_{\tq}$.
In (a)-(c) the gluino masses
are $m_{\tg} (m_{\tg}) =125~GeV$ (dashed), $200~GeV$ (solid),
$300~GeV$ (dot-dashed). Evolution from
$m_{\tq}$ is for $\Lambda ^{(4)} =300~MeV$ and the usual
quark mass thresholds.}
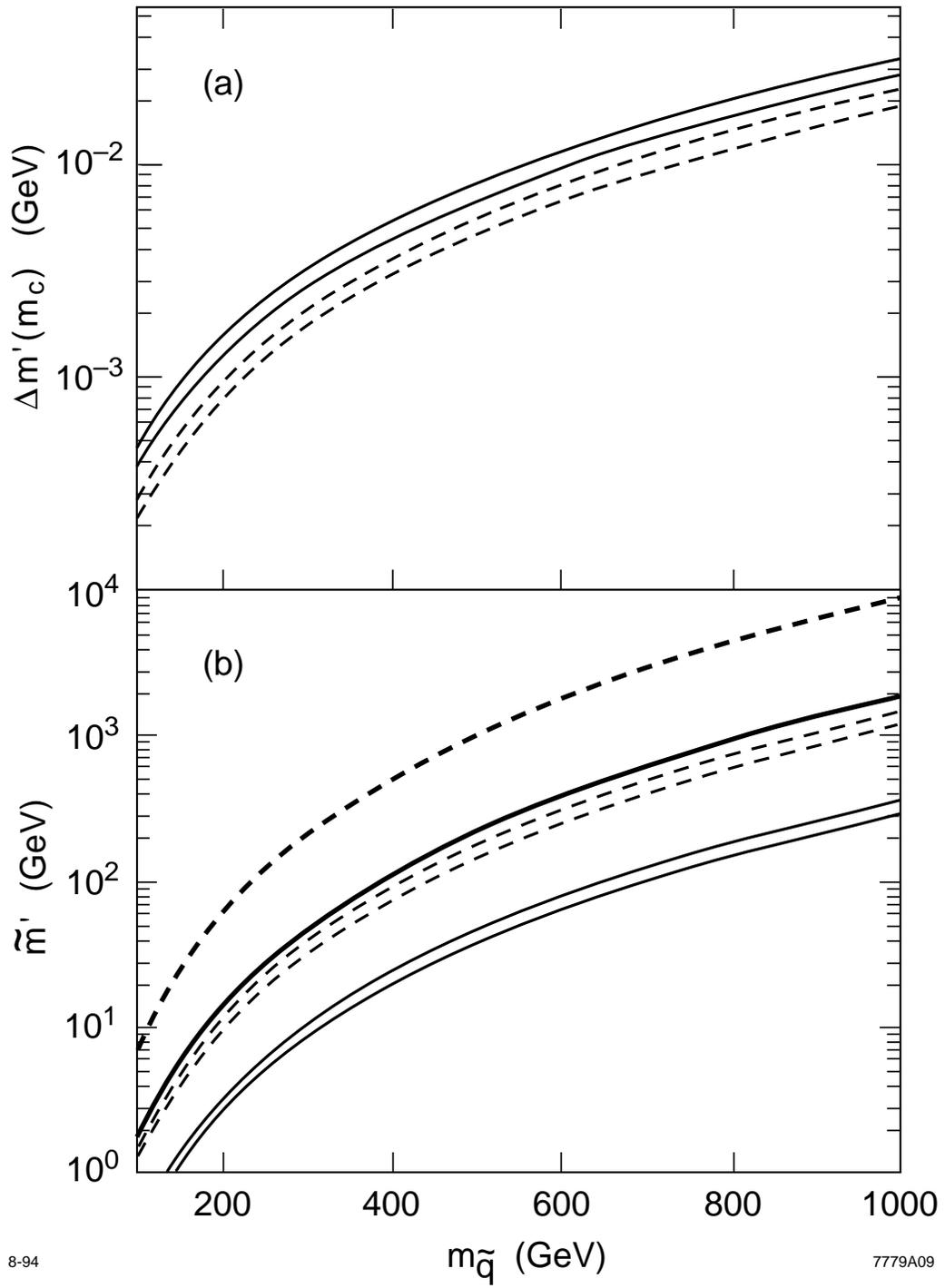
\item[Figure 11.] {Figs. 11a,b are the same as Figs. 10a,b for
$m_{\tg} = 4 ~GeV$ (solid) and $1~GeV$ (dashed).
Evolution from $m_{\tq}$  is for
$\alpha_s (M_z) = .124$, and the usual quark mass thresholds.}
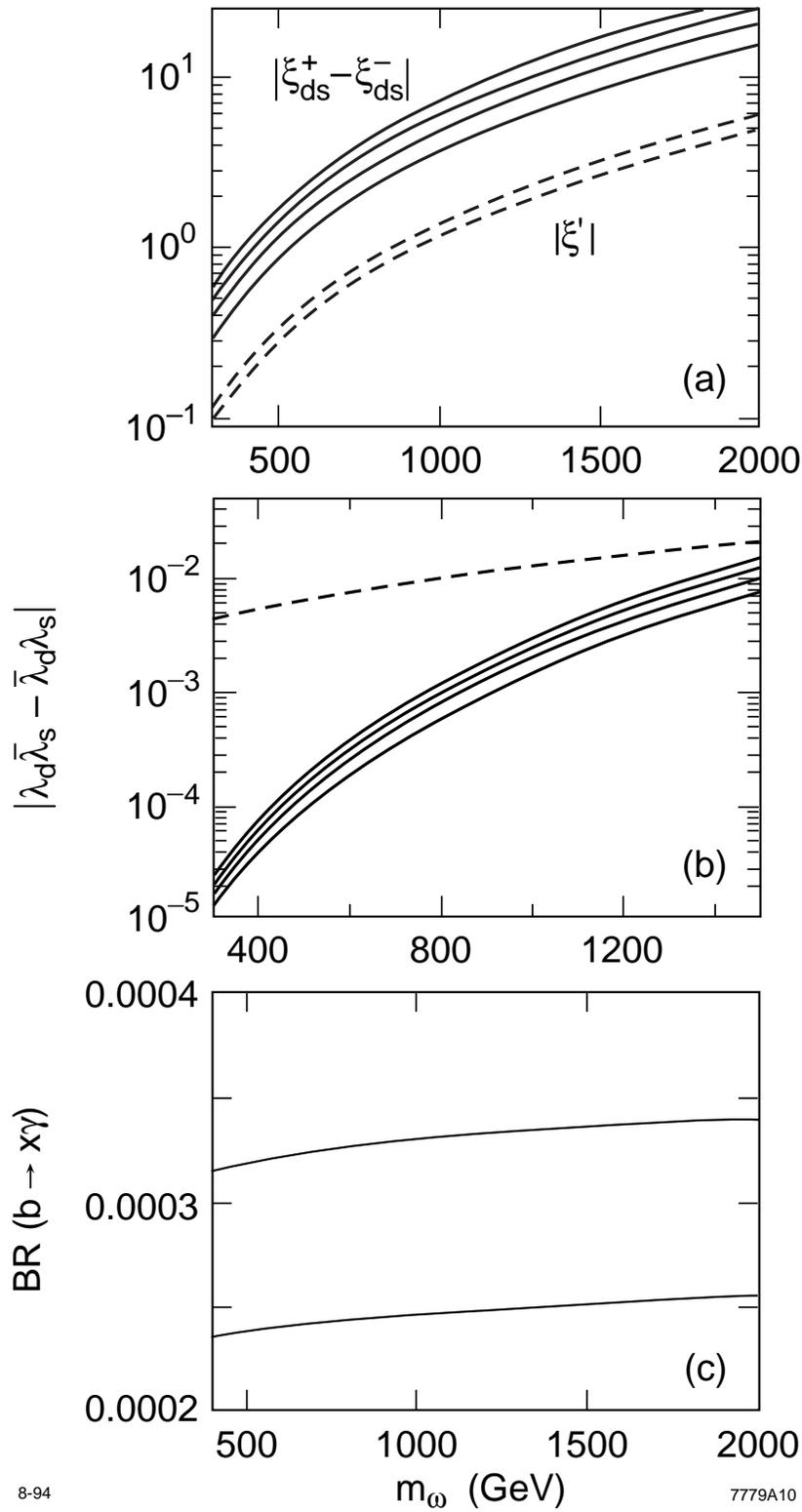
\item[Figure 12.] {(a) Contours of $R_0 =(1.5,1.25,1.0,.75){m_K^2 \over
\Lambda^2}$
(solid)
in the plane of $\xi_{ds}^+$ vs. $m_{\tq}$,
and contours of $\B=.10,.11$ (dashed) in the
plane of $\abs{\xi'}$ vs. $m_{\tq}$.  $\Lambda ^{(4)} =300~MeV$,
$m_t =170~GeV$, $m_b =4.8~GeV$, $m_c = 1.4~GeV$ and
$m_s (m_c) =150~MeV$.
(b) Contours of $R_0 = (1.5,1.25,1.0,.75){m_K^2 \over \Lambda^2}$ (solid)
and $\dmk =\dmk^{exp} $ (dashed) in the plane of
$\abs{\l_d \overline{\l}_s -\overline{\l}_d \l_s } $ vs.$m_{\om}$,
for $N_D =1$ and $N_{TC}=4$. (c) Contours of $\B =.10,.11$ in the
plane of ${\rm BR }(b \to x \gamma)$ vs. $m_{\om}$.  }
\end{itemize}

\end{document}